\documentclass[preprint]{aastex}

\shorttitle{SyXB System V934 Her}
\shortauthors{Hinkle et al.}

\begin{document}

\title{INFRARED SPECTROSCOPY OF SYMBIOTIC STARS. XII. \\ 
THE NEUTRON STAR SyXB SYSTEM 4U~1700+24 = V934 HERCULIS}

\author{KENNETH H. HINKLE}
\affil{National Optical Astronomy Observatory\\
P.O. Box 26732, Tucson, AZ 85726, USA}
\email{khinkle@noao.edu}

\author{FRANCIS C. FEKEL}
\affil{Tennessee State University, Center of Excellence in Information
Systems, \\
3500 John A. Merritt Boulevard, Box 9501, Nashville, TN 37209, USA}
\email{fekel@evans.tsuniv.edu}

\author{RICHARD R. JOYCE}
\affil{National Optical Astronomy Observatory,\\
P.O. Box 26732, Tucson, AZ 85726, USA}
\email{joyce@noao.edu}

\author{JOANNA MIKO{\L}AJEWSKA}
\affil{Nicolaus Copernicus Astronomical Center, Polish Academy of Sciences, \\
Bartycka 18, PL-00-716 Warsaw, Poland}

\author{CEZARY GA{\L}AN}
\affil{Nicolaus Copernicus Astronomical Center, Polish Academy of Sciences, \\
Bartycka 18, PL-00-716 Warsaw, Poland}

\author{THOMAS LEBZELTER}
\affil{University of Vienna, Department of Astrophysics\\
T\"urkenschanzstrasse 17, A-1180 Vienna, Austria}
\email{thomas.lebzelter@univie.ac.at}

\begin{abstract}
V934~Her = 4U~1700+24 is an M giant--neutron star (NS) X-ray symbiotic
(SyXB) system. Employing optical and infrared radial velocities spanning
29 years combined with the extensive velocities in the literature,
we compute the spectroscopic orbit of the M giant in that system.
We determine an orbital period of 4391 days or 12.0 yr, the longest for 
any SyXB, and far longer than the 404 day orbit commonly cited for this 
system in the literature.
In addition to the 12.0 yr orbital period we find a shorter period
of 420 days, similar to the one previously found. Instead of orbital
motion, we attribute this much shorter period to long secondary
pulsation of the M3~III SRb variable. Our new orbit supports earlier
work that concluded that the orbit is seen nearly pole on, which
is why X-ray pulsations associated with the NS have not been detected.
We estimate an orbital inclination of 11$\,.\!\!^\circ$3 $\pm$ 
0$\,.\!\!^\circ$4.  Arguments are
made that this low inclination supports a pulsation origin for the
420 day long secondary period. We also measure CNO and Fe peak
abundances of the M giant and find it to be slightly metal poor compared
to the Sun with no trace of the NS forming SN event.  Basic properties
of the M giant and NS are derived.  We discuss the possible
evolutionary paths that this system has taken to get to its current
state.  \end{abstract}


\section{INTRODUCTION}

Symbiotic X-ray binaries (SyXB) are a rare class of low-mass, hard
X-ray binaries consisting of a neutron star (NS) accreting mass
from an M giant \citep{murset_et_al_1997}.  The much more common
symbiotic systems (SySt) contain a white dwarf accreting mass from,
typically, a K or M giant.  SySt are identified by emission lines in
the optical that result from accretion processes.  The SyXB differ
from the SySt in having nearly normal optical spectra. Unlike SySt
that are found because of their optical emission lines, typical 
SyXB are first identified as X-ray sources and then later associated 
with M giant optical counterparts.

Since the companion to the NS is a low mass M giant, SyXB are also
classified as low-mass X-ray binaries (LMXB).  As described by
\citet{liu_et_al_2007}, typical LMXB have orbital periods of days
with the low-mass star transferring matter by Roche-lobe overflow
to the NS primary.  The low-mass star can be a white dwarf, a
main-sequence star, or an F-G subgiant.  SyXB differ from the larger
group of LMXB in having a giant companion to the NS, an orbital
period of years, and an exceedingly slow NS spin of minutes to hours
\citep{lu_et_al_2012, enoto_et_al_2014}.  To date, the total number
of confirmed SyXB systems is barely over a half dozen with the
Galactic population estimated to be $\sim$100--1000 \citep{lu_et_al_2012}.

While the NS must result from a supernova (SN), there are multiple 
possible evolutionary paths.  The companion star to the NS
in SyXB systems serves as a probe of the evolution of both objects.
The SN event that created the NS might seem to exclude the continued
presence of a stellar companion.  The ZAMS binary progenitors of
SyXB consisted of a massive star or massive stars with a low mass
companion.  The formation of the NS and the survival of the binary
have been widely discussed for all types of NS binaries.  
For core collapse supernovae (CCS) small asymmetries in the
explosion result in large velocities for the NS remnant and this
could easily disrupt the binary \citep{dewey_cordes_1987}.  To avoid
this problem other routes for forming the NS have been discussed,
for instance, the rotationally delayed, accretion-induced collapse
of a white dwarf \citep{freire_tauris_2014}.  Other schemes involve
common envelope phases.  For a high mass--low mass system a common
envelope stage could occur at the supergiant stage followed by the
explosion of the stripped, evolved supergiant core
\citep{taam_sandquist_2000}.  \citet{iben_tutukov_1999} discussed
a SyXB system resulting from triple system evolution.  For a massive
binary with a distant companion the massive binary could undergo
various merger, common envelope, and SN events.  However, population
synthesis calculations favor CCS
to create the NS \citep{lu_et_al_2012,zhu_et_al_2012}.

Understanding the known systems is an obvious prerequisite to sorting
out the evolutionary tracks.  There is little information about
either the cool star or the orbital parameters for most SyXB.  In
a previous paper of this series we undertook a detailed study of
the M~III in the bright X-ray SyXB, faint SySt system, GX1+4=V2116~Oph
\citep{hinkle_et_al_2006}.  
The GX1+4 system has an orbital period of 3.18 years with a NS spin period
of $\sim$2 minutes \citep{gonzalez-galan_et_al_2012}.  Here
we take a detailed look at optical and near-IR spectra of V934~Her
= HD~154791 = 4U~1700+24.  V934~Her is much less active than V2116~Oph,
which makes it a more typical example of the SyXB class.  Its spectrum 
is that of a typical early M giant with no optical emission lines 
\citep{goranskij_et_al_2012}. 
A peculiarity of this system is that
no periods have been found in the X-ray data, hence the spin period
of the NS is unknown.  \citet{galloway_et_al_2002} and \citet{masetti_et_al_2002} have attributed
this lack of an X-ray periodicity to the NS being seen close to
pole on.

We start by presenting a brief review of previous work on HD~154791.
We then discuss the extensive set of velocity observations.  Using
this data, we determine the orbital elements of the M giant and
discuss the contribution to the velocities from the stellar pulsation.  
This is followed by a section on stellar parameters for the M giant 
and an analysis of the stellar abundances.  Finally, we discuss 
the evolution of the M giant and the binary system.

\section{A BRIEF REVIEW OF HD 154791 = V934 Her = 4U 1700+24}

The X-ray source 4U 1700+24 was discovered roughly simultaneously
by \citet{cooke_et_al_1978} in Ariel V scans for high-latitude X-ray
sources and by \citet{forman_et_al_1978} from the Uhuru X-ray
catalog.  \citet{garcia_et_al_1983} found the Einstein X-ray
position to be coincident with the $V$ = 7.6 mag normal M giant
HD~154791.  Using standard models for stellar wind accretion,
\citet{garcia_et_al_1983} showed that a binary model with a NS
accreting mass from an M giant was a plausible explanation for the
X-ray luminosity and energy distribution.  Lack of velocity variations
$\gtrsim$5 km~s$^{-1}$ over an eight month period suggested either
that the system has a very long orbital period or that it was viewed
nearly face-on.

\citet{garcia_et_al_1983} found three emission lines in the IUE 
ultraviolet spectrum of HD~154791 that are not seen in normal M giant 
spectra. \citet{dal_fiume_et_al_1990} found that these UV emission lines 
have variable strengths associated with variations in the X-ray flux,
strengthening the connection with an accretion
process.  In addition, \citet{brown_et_al_1990} found the He~I 10830 \AA\
line is present with strong emission and absorption.  The He~I 10830
\AA\ 2 $^3$S - 2 $^3$P line has a metastable lower state 20 eV above
the ground state and is diagnostic of binary star X-ray activity.
However, \citet{sokoloski_et_al_2001} found no flickering in $B$ 
with a limit of $\sim$10 mmag.  

\citet{garcia_et_al_1983} identified the optical spectral type of 
HD~154791 as M3 II.  However, \citet{masetti_et_al_2002} found its 
spectrum to be a poor
fit to standard M3 II template spectra and preferred M2 III.  With
standard values for the bolometric magnitude of an M2 III a distance
of 420 $\pm$ 40 pc results, in good agreement with the $Hipparcos$
distance of 390 $\pm$ 130 pc.  The connection of the M giant and
the X-ray source was cemented by the \citet{masetti_et_al_2006}
measurement of a Chandra position for the X-ray source with an 
uncertainty of $\pm$0\farcs6, in
excellent agreement with the optical $Hipparcos$ position of the M giant.
From time series photometry provided by the $Hipparcos$ team
\citet{kazarovetsetal1999} assigned HD~154791 the variable star
name V934~Her.

\citet{masetti_et_al_2002} and \citet{galloway_et_al_2002} both 
noted that assuming a typical M2~III luminosity of 550~$L_\odot$, 
the M giant is about 200 times more luminous than the X-ray source.  
This explains the lack of
rapid optical variations since the contribution from
the X-ray source is negligible compared to the M giant
optical and UV flux. 
This also explains why the optical spectrum is
not peculiar.  In the case of the SyXB V2116 Oph/GX 1+4, a SyXB
with symbiotic emission lines, the stellar luminosity is four times
less than the X-ray luminosity.

4U~1700+24, the variable X-ray source component of the binary\footnote{In this paper we refer to the 
SyXB system observed in the optical and infrared as V934~Her and reserve the name 4U~1700+24 for the X-ray source.
However, as reflected in the title of this paper the optical and X-ray names 
are fully synonymous through most of the literature.}, does not have 
any periods detectable in the 2 to 2700 sec range \citep{garcia_et_al_1983}.
\citet{galloway_et_al_2002} similarly concluded that 4U 1700+24 is 
different from other
NSs detected in the X-ray region since no coherent or quasi-periodic
oscillations could be seen in the X-ray data.  \citet{masetti_et_al_2002}
confirmed that 4U 1700+24 has substantial X-ray variability but
this lacks periodicity.  This paper and \citet{galloway_et_al_2002} both
concluded that
the lack of periodicity results from viewing the NS nearly pole-on
with the magnetic axis aligned to the NS spin axis.  In this geometry,
hot spots on the NS will be continuously in view.

\citet{masetti_et_al_2002} found
that the size of the X-ray emitting area to be on the order
of tens of meters.  This suggests that the accretion is funneled
by the magnetic field onto the magnetic polar cap.
\citet{masetti_et_al_2002} noted that the presence of an M giant
wind was inferred from both the UV variability and the IRAS 12 and
25 $\mu$m measurements of a mid-IR excess.  An accretion
rate of $\sim$10$^{-14}$ $M_\odot$ yr$^{-1}$ was shown to be consistent with normal
values for both a red giant mass-loss rate, $\sim$10$^{-9}$ $M_\odot$
yr$^{-1}$, and accretion efficiency onto a NS of $\sim$10$^{-4}$.

\citet{galloway_et_al_2002} acquired high-resolution spectra of 
V934~Her in a 44~\AA\ region around 5200~\AA\ on 83 occasions.  Their
observations span nearly 15 years starting in 1982. A search for
periods in the 50--1000 day range found a 3.3 $\sigma$ period at about
410 days.  An elliptical orbit was then fit to the data resulting
in a 404 $\pm$ 3 day period. That orbit had a semi-amplitude of
0.75 $\pm$ 0.12 km s$^{-1}$ and an eccentricity 0.26. Given an
orbital period of $\sim$400 days and typical masses of 1.4 M$_\odot$
for the NS and 1.3 M$_\odot$ for the M giant, \citet{masetti_et_al_2002}
found from Kepler's third law a semi-major axis of
$\sim$300~R$_\odot$ and an orbital velocity of $\sim$30 km s$^{-1}$.
An inclination of $\lesssim$5$^\circ$
is required to match the observed velocity amplitude. 
The probability that a binary inclination
will be less than or equal to an inclination $i$ is $1~-~cos(i)$.  
An inclination of 5$^\circ$ or less has a probability of less than 0.5\%.

\citet{tiengo_et_al_2005} identified the O~VIII Ly-$\alpha$ line red
shifted by $\sim$3500 km s$^{-1}$ at 19.19 \AA\ in the
X-ray spectrum of 4U 1700+24.  They found this is in
agreement with the emitting gas being accreted by the NS at the
magnetospheric radius.  \citet{nucita_et_al_2014} found that this
is $\sim$1000 km above the NS. Again the requirement is that 
the system is observed nearly pole on with the magnetic and rotation
poles aligned.  The O~VIII line observation was confirmed by
\citet{nucita_et_al_2014}. They  
endorsed both the small size of the X-ray emitting area and the
nearly pole on aspect of the NS.  Both
\citet{gonzalez-galan_et_al_2012} and \citet{lu_et_al_2012} argued that
mass transfer in SyXB occurs through quasi-spherical wind-accretion
flowing along NS magnetic field lines.
Thus the SyXB differ from symbiotic binaries in not having accretion disks.
In the case of quasi-spherical accretion, rather than disk accretion,
the prominent optical emission features associated with disk 
accretion are not present.
\citet{krimm_et_al_2014} and \citet{burrows_et_al_2015} reported
on a series of X-ray flares observed by Swift where the radiation became 
harder as the luminosity increased.  
In agreement with the other models for 4U 1700+24 
the analysis 
requires an extremely compact accretor. 

\section{NEW OBSERVATIONS AND REDUCTIONS}

We observed the spectrum of V934~Her at high resolution in the optical 
and near-infrared on 90 occasions using five telescopes at four different
observatories and with six different instruments (Table~1).  The
extensive set of observations was made possible because V934~Her
is bright, $K$$\sim$3 mag, in the near-infrared  but not so bright as to
be unobservable with large telescopes.  The initial observation in
our data set was obtained in 1988 July.  However, our monitoring
of V934~Her started more than an decade later on 2000 July 13 when
we observed a section of its $H$~band spectrum with the Phoenix
cryogenic echelle spectrograph at the f/15 focus of the Kitt Peak
National Observatory (KPNO) 2.1\,m telescope.  The most recent set
of velocity observations, which have continued into 2017, were
acquired with the Fairborn 2~m telescope and fiber fed echelle
spectrograph.  Thus, our velocity data set spans 29 years.

The first observation reported here was obtained with the KPNO 4~m
telescope and Fourier Transform Spectrometer (FTS) on 1988 Jul 3
as part of a program to study abundances.  While FTS observations
are a gold standard free from systematics in both frequencies
and intensities, the technique suffers from multiplex disadvantage
and is best applied to bright stars \citep{ridgway_hinkle_1987}.
The spectrum covers the $K$ band at an apodized resolution, R =
$\lambda/\Delta\lambda$, of $\sim$32000.  The peak signal-to-noise
ratio is 63 and required a 70 min exposure.  The 4~m FTS is discussed
by \citet{hall_et_al_1978}, and the reduction techniques are discussed
by \citet{hinkle_et_al_1982}.  In addition to using the FTS observation
to determine a radial velocity, the spectrum was ratioed to a
spectrum of $\alpha$~Lyr that was observed on the same night.  We
analyzed this ratioed spectrum as part of our abundance analysis.
The FTS spectrum was also convolved to a resolution of 1.4 cm$^{-1}$
(R$\sim$3000) to compare it with \citet{wallace_hinkle_1997} spectra
of normal field stars.

The Phoenix data were acquired with either the KPNO 2.1 m or 4 m
telescopes or the Gemini South 8 m telescope.  A complete description
of the spectrograph can be found in \citet{hinkle_et_al_1998}.  The
Gemini South observation has the highest resolving power, R = 70000.
The other four Phoenix observations were taken with the widest slit
resulting in R = 50000.  Phoenix spectra cover a small, 0.5\%,
wavelength interval in several regions of the $H$ and $K$ band.

In 2000 October we observed a section of the $H$ band spectrum using
the KPNO 0.9 m coud\'e feed telescope and coud\'e spectrograph.
The detector was an infrared camera, NICMASS, developed at the
University of Massachusetts.  The 2-pixel resolving power is 44000
with the observation centered at 1.623 $\mu$m.  V934~Her was also
observed with the same detector, order sorting filter, and support
electronics at the Mount Stromlo Observatory (MSO) 1.88\,m telescope 
and coud\'e
spectrograph in 2001 and 2002.  In the MSO data the 2-pixel resolving
power is 24000.  The experimental setup that used the NICMASS camera
is described in \citet{joyce_et_al_1998} and \citet{fekel_et_al_2000}.
The Canberra area bush fires of 2003 January destroyed the MSO
1.88\,m telescope, spectrograph, and the NICMASS camera.

Following the loss of our equipment in Australia, we continued
observations at KPNO using the 0.9 m coud\'e feed telescope, coud\'e
spectrograph, and a CCD designated LB1A.  The 1980 $\times$ 800
pixel CCD was manufactured by Lawrence Berkeley National Laboratory
and is 300 $\mu$m thick. Our spectrograms, centered near 1.005
$\mu$m, have a wavelength range of 420 \AA\ and a resolving power
of $\sim$21500.  The coud\'e feed was closed as a result of NSF
budget cuts in 2014.

As noted above, unlike typical SySt the optical spectrum of V934
Her does not contain conspicuous emission lines or extensive veiling
caused by continuum emission.  As a result, for V934~Her it is
possible to acquire useful optical spectra and to measure radial
velocities of the M giant without complications. Thus, in 2015 February
observations were commenced with the Tennessee State University 2~m
Automatic Spectroscopic Telescope (AST) and fiber fed echelle
spectrograph \citep{ew2007}. The detector is a Fairchild 486 CCD
that has a 4096 $\times$ 4096 array of 15 $\mu$m pixels \citep{fetal2013}.
Forty eight echelle orders are covered ranging in wavelength from
3800--8260~\AA.  The observations were made with a fiber that
produces a resolving power of $\sim$25000 at 6000~\AA. 

For the near IR spectra standard observing and reduction techniques
were used \citep{joyce_1992}.  Wavelength calibration of Phoenix data,
KPNO coud\'e data, and Mount Stromlo coud\'e data posed a challenge because
the spectral coverage is too small to include a sufficient number
of ThAr emission lines for a dispersion solution.  Our approach was
to utilize absorption lines in a K giant to obtain a dispersion
solution.  Several sets of lines were tried, including CO, Fe~I,
and Ti~I.  These groups all gave consistent results.

Radial velocities of the program stars for the KPNO, MSO,
and Gemini South spectra were determined with the IRAF
cross-correlation program FXCOR \citep{fitzpatrick_1993}.  The
reference star was $\delta$~Oph, an M giant IAU velocity standard,
for which we adopted a radial velocity of $-$19.1 km~s$^{-1}$ from
the work of \citet{scarfe_et_al_1990}.

\citet{fetal2009} provide a general explanation of the velocity
measurement of AST spectra. In the particular case of V934~Her 
we selected a subset of 40 lines from our solar-type star line list that 
are relatively unblended in M giant spectra and range in wavelength
from 5000 to 6800~\AA.  Our unpublished velocities of several 
IAU radial velocity standards from spectra obtained with the 2~m AST 
have an average velocity difference of $-$0.6 km~s$^{-1}$ when 
compared to the results of \citet{scarfe_et_al_1990}. Thus, we 
have added 0.6 km~s$^{-1}$ to each of our AST velocities.

Figure 1 plots all our velocities as well as those from the 
Center for Astrophysics (CfA) that were provided by
D. Galloway (private communication 2017) and are discussed in 
Section 4.

On 2018 Apr 22 we obtained a spectrum of the $H$ and $K$ region of
V934~Her at R=45000 using IGRINS \citep{park_et_al_2014} on Gemini
South.  The integration time was a few seconds, so the spectrum
does not contain OH night sky lines for velocity calibration.  While
the wavelength/velocity calibration could be done using telluric
absorption lines, for the current paper we opted to use this spectrum
only for abundance analysis.  Since the spectrum has larger wavelength
coverage than even the archival FTS spectrum, it became a key element
in the abundance analysis.  We used the pipeline reduced IGRINS
spectrum, and to fit the continuum, we used the IRAF continuum
routine $splot~'t'$ at low order.  For our analysis it was necessary
to join the echelle orders to produce a $K$ band and an $H$ band
spectrum.  We did this by comparing the overlap regions between the
orders.  Our $H$ band analysis of this spectrum is based on the
1.5--1.7 $\mu$m region that is utilized by the APOGEE project
\citep{majewski_et_al_2016}.  Use of this region was facilitated
by the comprehensive line list developed by APOGEE \citep{She2015}.

In addition to the IGRINS spectrum we selected seven other spectra
for use in our chemical abundance analysis of V934~Her.  In Table~2
the observational details of the abundance analysis spectra are
provided.  An identifier (column 1) is given, which will be used
later when it is necessary to specify individual spectra.  We
analyzed the FTS spectrum since it covers the entire $K$ band roughly
20 years prior to the IGRINS observation.  However, both the S/N
and resolution are inferior to the IGRINS spectrum.  To supplement
these data we also included two $K$ band Phoenix spectra that cover
narrower ($\sim$100\,\AA) regions, one at 2.31 $\mu$m and a second
at 2.22 $\mu$m, and three $H$ band Phoenix spectra that cover a
narrow region ($\sim$65\,\AA) at $\sim1.56\mu$m.  For all the
abundance data a telluric reference spectrum of a hot star was
observed at approximately the same time.  With this reference
spectrum the telluric lines have been ratioed from the V934 Her
spectra.

\section{ORBITAL ELEMENTS}

The observed velocities (Fig.~1) suggest a long period orbit.
We searched for an orbital period in our radial velocity 
data using the least string 
method as implemented by T. Deeming \citep[PDFND,][]{betal1970}. 
Given the small amplitude of any orbital velocity variation 
plus the uncertainties of the 
velocities the possible periods cover a broad range 
from about 4200 to 4950 days 
with a best period at 4425 days.  This means that our 
extensive velocity time series (Fig.~1), aside from our initial 
FTS spectrum, covers just 1.4 orbital cycles.
With all our velocities given unit weight we obtained an orbital
solution with the SB1 orbit program \citep{betal1967}. Because
of the broad range of possible periods noted above, we tried
starting values of the orbital period from both the low end and
the high end of the 4200 to 4950 day range. In each case, the
orbit program converged to the same set of orbital elements resulting
in a period of 4479 days. 

While \citet{galloway_et_al_2002} discussed the Center for Astrophysics
spectra and velocities for V934~Her, individual velocities
were not published.  Fortunately, D. Galloway (private communication
2017) provided them to us. To check the compatibility of the
zero-points for our velocities and those from CfA, we compared the
orbital solution determined from our elements with the CfA velocities.
There was good agreement with the CfA velocities primarily being
distributed in the orbit at phases where there was little orbital
velocity variation.  After comparing the variances of the velocities
in the two orbital solutions we combined the two data sets, assigning
weights of 0.6 to the CfA velocities, and obtained a combined-data
solution for the orbital elements.  In the combined velocity solution
the orbital period decreased to 4394 days, about a 2~$\sigma$ change.
The eccentricity was likewise reduced by about 2~$\sigma$ with the
semi-amplitude increased by less than 1~$\sigma$.

We next looked at the velocity residuals from the combined data
orbital fit. A period search from 100--600 days with the program
PDFND was carried out on the CfA velocity residuals and resulted
in a best period of 406 days, similar to the value found by
\citet{galloway_et_al_2002}. We then made a separate period search
of the velocity residuals for our data. 
A period is clearly present in the data at greater than 10~$\sigma$ 
in the range 412 $\pm$ 10 days. 
Since both sets of velocities appear to have a second
periodicity of about 410 days, our last step was to analyze the two
sets of velocities with the general least squares (GLS) program of
\citet{d1966} to obtain a simultaneous solution for the short- and
long-period velocity variations. This final solution resulted in
periods of 420.2 $\pm$ 0.8 days and 4391 $\pm$ 33 days, respectively.
The uncertainties are 1~$\sigma$.

Table~1 provides the individual radial velocities for both the CfA
and our data. That table lists for each observation the heliocentric
Julian date, the observed total velocity, and the observed minus
calculated velocity residual ($O-C$) to the combined orbit. Also
computed and listed in the table are the long period orbital phase,
the long period velocity, which is equal to the total velocity minus
the computed short period velocity, the short period orbital phase,
and the short period velocity, which is equal to the total velocity
minus the computed long period velocity. The last column gives the
source of the observation. Table~3 provides the orbital elements
for both the short- and long-period variations. Although characterized
by orbital parameters, the short-period variations, as will be 
discussed later, result from long secondary period (LSP) velocity 
changes rather than a third component of the system. The very small 
value of the long-period
orbit mass function, 0.0022 $\pm$ 0.0005 $M_{\sun}$, suggests that 
our 4391 day orbit is seen nearly pole on. We will return to this 
point when defining the stellar parameters. 

Figure~2 presents the computed velocity curve of the long-period
orbit compared with the radial velocities, where zero phase is a
time of periastron.  Each plotted velocity consists of the total
observed velocity minus its calculated short-period velocity.
Figure~3 shows the computed velocity curve of the short-period 
``orbit'' compared with the KPNO radial velocities, where zero 
phase is a time of periastron.  Each plotted velocity consists 
of the total observed velocity minus its calculated long-period 
velocity.

\section{STELLAR PARAMETERS}

\subsection{Photometric Periods}

\citet{tomasella_et_al_1997} acquired $UBVRI$ photometry on six
nights over a two month period and found no variability at $V$ and
$B$, although the values were a few 0.1 mag different from
those previously reported by \citet{garcia_et_al_1983}.
$Hipparcos$ found that V934~Her varied by 0.16 mag with a possible
period of 31 days.  This forms the basis of the General Catalogue of
Variable Stars SRb designation \citep{kazarovetsetal1999}.  
\citet{goranskij_et_al_2012}, using precision photometry, found periods of 28, 31, and 44 days
in $V$, 29, 44 and 405 days in $B$, and 44 and 415 days in $U$.  
Semi-regular variables characteristically have simultaneously excited
closely separated periods from the same overtone \citep{hartig_et_al_2014}.
The  amplitudes
in $B$ and $V$ are $\sim$0.05 mag, so small as to be easily missed
by earlier work.  Similarly \citet{gromadzki_et_al_2013} found a
period of 44 days.  

\subsection{The 400 Day Period}

A common characteristic of SR variables is a long secondary period
(LSP) to the dominant pulsation period.  The LSP is typically 8--10
times longer than the dominant period \citep{nicholls_et_al_2009,
hartig_et_al_2014}.  Taking the
V934~Her photometric period to be 28--44 days \citep{goranskij_et_al_2012,
gromadzki_et_al_2013}, then the LSP is the $\sim$400 day period.

In M giants LSPs can be detected in both luminosity and velocity
variations.  As noted above, the first orbit for V934~Her was based
on the \citet{galloway_et_al_2002} radial velocity period of 404 $\pm$ 3
days.  If the LSP velocity variations are interpreted as an orbit,
the velocity curve is distinctive with $\omega$ $\sim$250$^\circ$
and $e$ $\sim$0.35 \citep{hinkle_et_al_2002}.  These parameters are
a reasonable match to the ``orbital'' elements of V934~Her presented
by \citet{galloway_et_al_2002}.  As discussed earlier, we have
computed a short period ``orbit'' with $\omega$ = 237$^\circ$ and
$e$ = 0.33.  We also note the similarity of these numbers to the
LSP ``orbit'', $\omega$ = 229.5$^\circ$ e = 0.33, of the very well
studied SySt CH~Cyg \citep{hinkle_et_al_2009}.

V934~Her presents an interesting case of LSP because the orientation
of the star is known.  Assuming that the rotation axis of the M
giant is parallel to that of the orbit, the star is seen nearly
pole on.  In this case models for the LSP that require semidetached
binaries \citep{wood_et_al_1999,soszynski_2007} and rotating spots
with dust formation \citep{takayama_et_al_2015} can be excluded.
As discussed by \citet{stothers_2010} and \citet{saio_et_al_2015},
this narrows the explanations to pulsation mechanisms involving
convection.  A global pulsation mechanism for LSP now appears to
be widely accepted if not fully understood \citep{trabucchi_et_al_2017}.

LSPs are associated with increased mid-IR excess
\citep{wood_nicholls_2009}.  In the case of V934~Her this is in
agreement with the results of \citet{masetti_et_al_2002} who found
a larger than expected IR excess.  \citet{masetti_et_al_2002}
reported a tentative period of $\sim$400 days from RXTE ASM
observations.  \citet{galloway_et_al_2002} analyzed the same data
extended by an additional year and refined this as a period of 404
$\pm$ 20 days.  The existence of the LSP in the X-ray data would
link the LSP to cyclic enhancements of mass loss from the M III.
\citet{corbet_et_al_2008} analyzed Swift BAT observations 
and RXTE ASM observations   
including data previously analyzed by
\citet{masetti_et_al_2002} and \citet{galloway_et_al_2002} 
but
was not able to find the $\sim$400 day period.

\subsection{Temperature, Luminosity, Surface Gravity}

\citet{garcia_et_al_1983} found that V934~Her had an optical spectral
type of M3 II while \citet{masetti_et_al_2002} determined an optical
spectral type of M2~III, which was confirmed by 
\citet{goranskij_et_al_2012}. Their photometry of V934~Her does not 
show any measurable reddening.
While \citet{GauPol1999} claimed the spectrum is abnormal, this
has been refuted \citep[see for instance][]{masetti_et_al_2002}. 
Other than the claim of \citet{GauPol1999}, there is no evidence for 
spectral variability. \citet{tomasella_et_al_1997} were not able to 
detect changes in the optical spectrum of V934~Her during a strong X-ray
outburst.

The FTS spectrum of V934~Her discussed earlier covers the 
2.0--2.5 $\mu$m near-IR $K$ band.  After apodizing to R$\sim$3000 
this spectrum was compared (Fig.~4) to M giant standards
from \citet{wallace_hinkle_1997}. The strong CO features mark V934
Her as a luminous star.  For the mid-M temperature classes a good
indicator of temperature is the Sc I 4600 cm$^{-1}$ line.  In V934~Her 
this line is approximately intermediate in strength between the
M2~III and M4~III spectra.  Other atomic features are also stronger
than in the M2 spectrum.  We assign a temperature classification of
M3.  Importantly, the infrared spectrum of V934~Her looks like a
normal star with no emission features in its $K$ band spectrum.

The distance to V934~Her has been discussed by both
\citet{garcia_et_al_1983} and \citet{masetti_et_al_2002} with 
their results differing by a factor of two.  This discrepancy 
has been resolved by the 
$Gaia$ parallax of 1.837 $\pm$ 0.032 mas, i.e. 
a distance of 544~$^{+10}_{-9}$ pc \citep{Gai2018}.  Combining the distance
with the galactic coordinates, V934~Her is 
296 pc above the galactic plane.
\citet{goranskij_et_al_2012} suggested that V934~Her is unreddened.  
Confirmation is provided by the images of \citet{schlafly_finkbeiner_2011}
who find E(B-V) is at most 0.038.  Ignoring reddening,
the 2MASS \citep{cutri_et_al_2003} $m_K$ = 2.988 mag results in an absolute $K$
magnitude $M_K = -5.690$ mag. Taking $J-K$ = 1.181 mag, the 2MASS color of
V934~Her, the $K$ band bolometric correction from 
\citet{bessell_wood_1984} is 2.92 mag. The resulting bolometric 
magnitude for V934~Her is $-2.775$ mag corresponding to 1028 $\pm$ 40~$L_\odot$ 
where the formal uncertainty is from the distance.  The uncertainties associated
with  the infrared photometry and bolometric correction are not available.

\citet{van_belle_et_al_1999} gives an effective temperature for an
M3~III of 3573 $\pm$ 22 K.  Alternately, using the $V-K$ color of
V934~Her, the $V-K$ color--$T_{eff}$ relation of 
\citet{van_belle_et_al_1999} yields 3677 K. 
\citet{dyck_et_al_1996}
suggests an
effective temperature for an M3~III 
of 3650 K.  Adopting a 3650 K effective temperature as a mean value, 
the literature photometry for V934~Her/4U~1700+24 is shown
in Figure~5, fit with a 3650 $K$ blackbody. 
The blackbody integrated flux is
1.4 $\pm$ 0.1 $\times$ 10$^{-10}$ W m$^{-2}$.  Correcting for the $Gaia$ distance
of 544 pc the
bolometric magnitude is $-3.1$ $\pm$ 0.1, i.e. $L = 1367$ $\pm$ 120 $L_\odot$.

We adopt mean values for the temperature and luminosity with
uncertainties embracing the range of values, $L = 1200$ $\pm$ 
200 $L_\odot$ and $T_{eff}$ = 3650 $\pm$ 100 K.  The values for the 
temperature and luminosity are in good agreement with both the observational HRD and 
the evolutionary tracks for an M3~III resulting from a 
low mass progenitor \citep{escorza_et_al_2017}.  Similarly, using 
the $Gaia$ distance, $K_S$, and $J-K$ colors the relations of 
\citet{lebzelter_et_al_2018} confirm that V934~Her is on either 
the RGB or faint AGB.

We have argued that the $\sim$410 day period of V934~Her is not
an orbital period but is a pulsational LSP. The Period-Luminosity 
relation of \citet{wood_2000} can be applied to the photometric periods.
\citet{goranskij_et_al_2012} and \citet{gromadzki_et_al_2013} found
periods of 28 and 44 days with an LSP of 410 days. The LSP is associated
with a primary period on the first overtone B sequence
\citep{wood_et_al_1999,trabucchi_et_al_2017}. 
We assume that the 44 day
period is the first overtone, B sequence period and the 28 day
period is the second overtone, A sequence period.  From the mid-line
of the relations for 28 day, 44 day, and 410 day periods the
corresponding LMC W$_{JK}$ from Figure~1 of \citet{trabucchi_et_al_2017}
or Figure~2 of \citet{soszynski_et_al_2007} is 11.37.  Assuming a
distance modulus of $m-M$ = 18.5 mag for the LMC, this corresponds 
to an $M_K = -6.3$, 0.6 mag brighter than measured.  However, the
$P-L$ relations have a width $>$0.5 mag in W$_{JK}$ so the $P-L$
bolometric magnitude is in agreement with the absolute $K$ mag
determined from the $Gaia$ distance.

The blackbody fit to the photometry yields the stellar radius as
well as the flux.  The uniformly illuminated radius required for
the blackbody is $91~^{+14}_{-21}$ $R_\odot$.  The \citet{bourges_et_al_2014}
data base gives a limb-darkened angular diameter computed from the 
colors of V934~Her of 2$R = 1.544$ $\pm$ 0.121 mas.  Using the $Gaia$
distance the red giant radius is 90 $R_\odot$.
\citet{van_belle_et_al_1999} give a smaller radius of 71  $R_\odot$
but the relationship has considerable width.

\subsection{Mass}

From the models of \citet{charbonnel_et_al_1996}
the luminosity of 1200 $L_\odot$ and $T_{eff}$ of 3650 K place 
V934~Her on the early AGB of a solar metallicity 1.7 $M_\odot$ star.  
STAREVOL tracks by \citet{escorza_et_al_2017} suggest a 
mass a few 0.1 $M_\odot$ smaller.  The NS companion in the V934~Her system has a limited range of mass.  
The upper limit to the mass of a NS occurs at $\sim$$3 M_\odot$ 
when the internal sound speed reaches the speed of light.  Such a 
large mass for the NS seems unlikely. Masses of NSs
in binary radio-pulsar systems are all very close to 1.35 $M_\odot$
\citep{thorsett_chakrabarty_1999}.  Masses larger than 1.35 $M_\odot$
might occur \citep{lorimer_mclaughlin_2006} but masses measured for LMXB NSs, 
which can be uncertain, 
seldom exceed 1.5 $M_\odot$ \citep{casares_et_al_2017}.

The mass function from the orbit of the M giant is $$ f(m) ~ = ~ {
M_{NS}^3 sin^3(i) } / { ( M_{RG} + M_{NS} )^2 }  = 0.0022 $$ If
we assume that $M_{NS}$=1.35 $M_\odot$ and $M_{RG}$=1.7 M$_\odot$ the
orbital inclination is 11$\,.\!\!^\circ$7.  Lower masses for the red
giant from different evolution models or mass loss drive the inclination smaller.  If,
as suggested by \citet{lu_et_al_2012}, the NS has accreted
mass from the giant, the inclination is also smaller.  For example,
an M giant mass of 1.4 $M_\odot$ reduces the inclination to 10$\,.\!\!^\circ$9.
We conclude that an orbital inclination in the range of 
11$\,.\!\!^\circ$3 $\pm$ 0$\,.\!\!^\circ$4 is in agreement with the mass estimates. 
The probability of an inclination of 11$\,.\!\!^\circ$7 or less is 2 percent.

If we adopt 
a mass for the M giant of 1.6 $M_\odot$, averaged between the evolutionary models, and 
a radius of 90 $R_\odot$ 
than the surface gravity is 5.4 cm s$^{-1}$, log $g$ = 0.7,
with the uncertainty in the mass and radius resulting in a uncertainly
in log g of $\sim$ 0.2.  
The parameters for the M giant are presented in Table 4.
For abundance determinations (Section 6) 
we have adopted atmospheric
models within the grid of model atmospheres of
$T_{\rm{eff}}=3650$\,K and $\log g= 0.5$.  

\section{ABUNDANCES}\label{abunds_sect}

\subsection{Methods}

Abundances were measured with the spectral synthesis technique
in the classical way, i.e., employing local thermodynamic equilibrium
(LTE) analysis based on a 1D, hydrostatic model atmospheres
\citep[MARCS, ][]{Gus2008}.  Synthetic spectra were calculated with the 
code developed by M.  Schmidt \citep[WIDMO, ][]{Sch2006}.
The general characteristics of the adopted method together with
its justification is discussed in a series of papers on chemical
composition analysis in SySt giants \citep[][and references
therein]{Gal2016}.  In summary the abundance
calculations for given model
atmospheres were performed as follows. The initial--starting
values for the free parameters were obtained by adjusting roughly
by eye the synthetic to the observed spectrum through several
iterations. Next, the simplex algorithm \citep{Bra1998} was used
for $\chi^2$ minimization in the parameter space. Besides the relevant
abundances and isotopic ratios, additional free parameters were the
line broadening for each spectrum expressed as a macroturbulent velocity,
$\zeta_{\rm t}$, and a microturbulent velocity, $\xi_{\rm t}$.  For
the V934~Her analysis $\xi_{\rm t}$ was found by examining the large 
range of excitation potentials and line strengths, especially from
$^{12}$C$^{16}$O lines over the broad wavelength range of the IGRINS spectrum.

The excitation potentials and gf-values for transitions in the case
of atomic lines in the narrow $H$-band region of the Phoenix spectra
were taken from the list by \citet{MalBar1999} and for the $K$-band
region from the Vienna Atomic Line Database \citep[VALD, ][]{Kup1999}.
For the molecular data in the $K$-band region we used line lists
by \citet{Goo1994} for CO, \citet{Kur1999} for OH, and 
\citet{Sne2014} for CN.  For the $H$-band IGRINS spectrum we used
the DR12 release of the APOGEE line lists \citep{She2015}. 

The spectrum synthesis was run with model stellar atmospheres
covering a broad range of effective temperature from 2900 to 4250\,K,
surface gravity from $0.0$ to $+1.0$, and metallicity from $-0.5$ to 
$0.0$. The data sets were fit separately since the data covered a range of
resolution and signal-to-noise ratio.  The regions of the spectra 
contaminated with artifacts
or with insufficiently well-reduced telluric absorption features
were excluded from the analysis.

\subsection{Limitations of the Model Atmosphere}

The parameters derived above for V934~Her, $T_{eff}$=3650 K, log g
= +0.5, and approximate solar metallicity, resulted in synthetic
spectra that were excellent fits to the $H$ band spectra (Fig.~6).  
However, to our surprise, the strong lines in the $K$-band region 
were best fit with a significantly lower effective temperature, 
$T_{\rm{eff}}=3100$\,K (FTS spectrum) and $T_{\rm{eff}}=3000$\,K 
(IGRINS spectrum).  The surface gravity remained $\log g = 0.5$ in 
all cases.  The best example of the poor fit by the 3650 K synthetic 
spectrum is for the CO first overtone where lines of different 
excitation potentials and strengths are present (Fig.~7).  The 
IGRINS spectra are especially interesting since the first and 
second overtone CO regions were observed simultaneously.  The 
$H$-band second overtone CO lines are fit by $T_{eff}$=3650 K while 
the first overtone CO lines in $K$-band region appear to require 
a $\sim$600 K lower temperature.

As noted IGRINS $K$ and $H$ spectra were taken simultaneously, hence,
explanations for the lower excitation temperature that invoke time
variability can be ruled out.  Since weak lines are fit by a 3650~K 
effective temperature while strong lines are not, the outer layers
of the model atmosphere must be too hot.  To further investigate
this problem we did a curve-of-growth analysis of the CO lines. This
technique, discussed by \citet{hinkle_et_al_2016}, requires
large spectral coverage, which was made possible by our IGRINS
observation.  Using the CO second overtone lines, we found a CO
excitation temperature of 3375~K.  The CO second overtone lines are 
generally weak, at most $\sim$30\% deep.  Comparison to a similar 
analysis of spectral standard M giants shows that a 3375~K 
excitation temperature corresponds to an effective temperature of 
$\sim$ 3650 K. This provides further confirmation that the spectral 
type has been correctly assigned.  On the other hand, the strong 
CO first overtone lines have a much lower excitation temperature.  
The relatively small sample of strong lines does not allow a 
solution but the excitation temperature is less than 3000~K.

The $K$ band region of the V934~Her spectrum contains measurable
lines from the CO isotopologues $^{13}$C$^{16}$O and $^{12}$C$^{18}$O.
These lines are not nearly as strong as the $^{12}$C$^{16}$O 2-0
lines. There are also clear upper limits for 2-0 $^{12}$C$^{17}$O
lines.  With $T_{exc}$=3375~K curves-of-growth
were computed for the isotopologues.  Shifts between these
curves-of-growth give the isotopic abundances (Fig.~8).  We find
$^{12}$C/$^{13}$C = 10 $\pm$ 4, $^{16}$O/$^{17}$O = 2500$^{+1500}_{-1000}$,
and $^{16}$O/$^{18}$O = 262 $\pm$ 100.  These values match the
values found from the spectrum synthesis.  The curve-of-growth
analysis compares weak second overtone $H$-band $^{12}$C$^{16}$O
with similar strength isotopic lines in the $K$-band.   Spectrum
synthesis uses a model atmosphere to fit a spectral interval.  The
failure of the synthetic spectrum to fit the strong lines must not be a
problem related to wavelength since the model works for both H and
K weak lines.  We also measured the CO lines in the $K$ band FTS
spectrum.  The equivalent widths from the FTS and IGRINS spectra
are in reasonable agreement, again demonstrating that there is not
a time dependent problem.

\citet{tsuji_1988} reports similar difficulties in fitting the CO first
overtone lines.  Tsuji found that extra absorption in low excitation
first overtone CO lines is a common property of late-type spectra.
He attributed the low temperature absorption to a quasi-static, turbulent,
1000--2000~K extended region in the outer atmospheres of these stars.  

\citet{chakrabarty_roche_1997} suggested that the NS in
the SyXB V2116~Oph system heats the red giant, altering the TiO band
strengths and impacting estimates of the spectral class based on
this molecule.  The orbit of V2116 Oph is close to edge on.  
\citet{chakrabarty_roche_1997}
derived a mean spectral class of M5.  In our analysis of V2116 Oph
\citep{hinkle_et_al_2006} we found that the M5~III effective
temperature, $T_{eff}$ = 3400~K, agreed with the effective temperature
determined from spectral synthesis of the infrared spectrum.  This
analysis was based on Phoenix spectra covering small regions of the
spectrum.  In 2018 April we observed V2116 Oph with IGRINS.
Using this observation, we obtained $T_{exc}$ = 3200~K for the CO 
second overtone.  This corresponds to $T_{eff}$ = 3370~K, so it is 
in good agreement the \citet{chakrabarty_roche_1997} spectral type.  
The separation of the NS and M giant in V934~Her is about two times 
larger than it is in V2116~Oph, so NS heating should be even less 
in V934~Her.

\subsection{Abundance results}

Table~5 lists final values of abundances obtained from the spectra
for $T_{\rm{eff}}$ = 3650\,K and log~$g$ = 0.5.  Resulting values for
the broadening parameters are presented in Table~6.  The contributions
to the uncertainties in the abundances are given in Table~7.
Uncertainties in the abundances come mainly from uncertainties in 
stellar parameters.  The final uncertainty in Table~7 is the 
quadrature sum of uncertainties of each model parameter.  IGRINS 
results derived entirely from the $H$-band are similar to FTS and 
Phoenix results from the $K$-band.  In spite of the difficulties in 
fitting the spectra with a consistent model atmosphere, the $K$- and 
$H$-band results (Table~5) are similar with the exception of the 
N abundance derived from the FTS spectrum.  We attribute this to the 
lower quality of that spectrum.  However, to err on the side of 
caution, in the subsequent discussion we use only the $H$-band 
results with the exception of the C and O isotopes.

\section{DISCUSSION}

\subsection{Stellar Evolution} 

The probability of forming a binary
system outside of a globular cluster by gravitational capture is
nearly zero.  
Stellar evolutionary tracks show that the main sequence mass of the
V934~Her giant was in the range $\sim$ 1.4--1.7 $M_\odot$, 
so the unevolved system was a binary consisting 
of the massive progenitor of the NS and a $\sim$1.6 $M_\odot$ companion.  
The $^{16}$O/$^{17}$O oxygen isotope ratio is
very large, $\ge$2000. 
This large value indicates that the ZAMS mass of
the M giant progenitor was low, $\lesssim$ 1.5 M$_\odot$ \citep{smith_1990,
hinkle_et_al_2016}.
The agreement of masses from the evolutionary tracks 
and abundances requires that mass transfer from the
proto-SN supergiant to the current M giant, if any, was no more than
a few 0.1 M$_\odot$.  
The main sequence 
lifetime for a 1.6 M$_\odot$ star is
$\sim$2 Gyrs \citep{charbonnel_et_al_1996} while the lifetime
of a $>$8 M$_\odot$ star is $\lesssim$100 Myrs
\citep{vassiliadis_wood_1993, tauris_van_den_heuvel_2006}.   Thus
the age of the NS is $\sim$2 Gyr.

The carbon $^{12}$C and nitrogen $^{14}$N abundances of the M III (Table~5)
reflect mixing during the first dredge-up.
This is confirmed by the low carbon isotope ratio, $^{12}$C/$^{13}$C
$\sim 7$ -- $11$ and is consistent with a red giant or early AGB status
for the M III.   
The giant in V934\,Her has a slightly
sub-solar metallicity.  Following \citet{lambert_1987}, we have
computed [$\alpha$/Fe] =  $+$0.27 from the average of the Mg, Si,
and Ca abundances.  The [$\alpha$/Fe] versus [Fe/H] is close to the
mean relation \citep{lambert_1987} and shows no notable peculiarity
for this star.  The $\alpha$ element and Fe abundances are similar
to the abundances of many other SySt giants \citep{Gal2016, Gal2017}.
\citet{casares_et_al_2017} report that in LMXB many of the low-mass
stars show enhancements of Fe and $\alpha$ elements. Modeling
suggests that this results from the capture of SN ejecta by the dwarf
companion.  We conclude that in the SyXB giants any SN ejecta on the surface has been
mixed into the interior as the star evolved up the giant branch.

\citet{lu_et_al_2012} discussed Monte Carlo simulations of the SyXB
population including  CCS, electron-capture SN (ECS), and
accretion-induced collapses (AIC).  Their study concludes that between
70\% and 98\% of SyXB NSs are formed via core collapse with the remainder
formed via ECS.  The simulation finds that systems forming via ECS have short
initial periods and have passed through a common envelope phase.  Similar 
scenarios are discussed by \citet{willems_kolb_2002}.

The simulated SyXB population of \citet{lu_et_al_2012} has typical parameters similar
to V934~Her.  Perhaps the existence of binary systems that survived a CCS 
should not be surprising since all massive
stars have at least one companion \citep{duchene_kraus_2013} and
the number of binary survivors of a SN is very small.
In the case of V934~Her the $Gaia$ proper motions of $-$10.06 $\pm$ 0.04 mas yr$^{-1}$ in
right ascension and $-$6.40 $\pm$ 0.05 mas yr$^{-1}$ in declination
correspond to velocities of $-$25.9 and $-$16.5 km s$^{-1}$ at the $Gaia$
distance of 544 pc. The $\gamma$ velocity for the binary is
$-$47.36 $\pm$ 0.06 km~s$^{-1}$ so the space velocity of this star is not
unusually large. 

Assuming a NS mass of 1.35 M$_\odot$, a mass of 1.6 M$_\odot$ for
the M giant, and an orbital period of 12.0 years, Kepler's third
law gives a semimajor axis $a$ of 7.52 AU for V934 Her.  At periastron the 
separation is $a(1-e)$ = 4.86 AU. From the formula of  
\citet{eggleton_1983} the M giant Roche lobe at periastron is 
1.93 AU or 415~$R_\odot$, which is much larger than the current 
stellar radius of $\sim$90 $R_\odot$. 
As a tip AGB star the stellar radius will increase to $\sim$250
$R_\odot$ \citep{ohnaka_et_al_2006}.  At the same time the mass-loss
rate will increase from the current $\sim$10$^{-9}$ $M_\odot$ yr$^{-1}$ to 
$\sim$10$^{-6}$ $M_\odot$ yr$^{-1}$.  Over the 10$^6$ yr TP AGB
lifetime \citep{vassiliadis_wood_1993} the current 1.6
$M_\odot$ M III will lose $\sim$1 M$_\odot$ to become a 0.6
$M_\odot$ white dwarf \citep{si_et_al_2018}.  As the mass loss increases,
mass transfer to the NS will decrease the orbital separation.  Simulations
by \citet{wiktorowicz_et_al_2017} predict the evolution of ultra-luminous X-ray
(ULX) sources from NS--low mass star binaries with masses
nearly identical to those in the V934~Her system. A complicating factor
is the increased absorption of the X-ray flux due to the 10$^{3}$ increase in mass loss\footnote{Noted
by the anonymous referee}.

\subsection{Orbital evolution}

From radial velocity observations \citet{fetal09} obtained
orbits for non-symbiotic M giant stars in the {\it Hipparcos}
survey and combined the results with M giant orbits from the
literature to produce a sample of 29 systems.
In a follow-up paper \citet{jetal09} examined the ($e$--log$P$)
diagram of those 29 M giant binaries from \citet{fetal09}.  Although
V934~Her has a degenerate companion it has an unremarkable optical spectrum. 
Thus, we compare V934~Her with the M giant sample of
\citet{fetal09}.

Figure~1 of \citet{jetal09} shows that M giants with periods
up to about 1500 days all have eccentricities below 0.25.
For M giants with longer period orbits, except for one nearly
circular orbit system, the eccentricities range from about
0.3 to 0.75. V934~Her with its period of 4391 days and
eccentricity of 0.35 clearly has a very non-circular orbit but is 
situated near the lower end of the eccentricity
distribution.
The kick velocity resulting from asymmetry during a CCS 
can be substantial to the point of disrupting the binary
\citep{lyne_lorimer_1994}.  Given the large, 4.8 AU, periastron separation
for V934 Her, 
tidal forces have not substantially acted to circularize the orbit.  The
eccentricity near the lower bound of M giants
may well reflect the primordial eccentricity of the system
and apparently was not significantly increased as a result of
the SN event.  
This suggests that the NS resulted from an ECS 
that has low kick velocity \citep{lu_et_al_2012}.

\subsection{Long Secondary Period}

The presence of a LSP pulsation of the M giant is supported by 
both spectroscopy and photometry.  The 420 day spectroscopic ``orbit''
is a close match to the velocity variations observed in
other LSP variables \citep{hinkle_et_al_2002}.
The luminosity derived from $Gaia$ places the 44 day photometric
period of V934~Her on the AGB pulsation first overtone, the 28 day 
period on the second overtone, and the 404  day
photometric period on the 
LSP sequence \citep{trabucchi_et_al_2017}.  Assuming the
stellar equator is aligned with the plane of the orbit, the M giant is seen
nearly pole on. This narrows the list of possible LSP mechanisms
to those favoring convection \citep{trabucchi_et_al_2017}.  Strong
absorption 
lines in the M giant spectrum are not well fit by a standard model 
atmosphere. A connection between the atmospheric structure and LSP 
is an area for future investigation.

Published observations show a tentative connection between the LSP and X-ray
activity, presumably driven by changes in the mass loss.   In the
SyXB/StSt system V2116~Oph/GX~1+4 activity is enhanced near periastron
passage \citep{ilkiewicz_et_al_2017}.  Although the V934 Her orbit
is significantly more eccentric than that of V2116~Oph, the periastron
separation, 2.28 AU, is still about twice that of V2116 Oph/GX 1+4.
It would be interesting to confirm the connection between 
LSP and X-ray activity and to see if the activity of V934 Her also
increases near periastron.

\section{CONCLUSIONS}

The NS--M giant symbiotic binary V934~Her is shown to have a 12 year orbit
with an eccentricity of 0.35.  The period previously found in the
velocity data, 404 days and revised here to 420 days, is not the
binary orbit but the LSP pulsation of the M giant. We find the M
giant to have a spectral type of M3~III and to have slightly subsolar
abundances.  The $^{16}$O/$^{17}$O is consistent with a progenitor main sequence
star having a mass similar to that determined from the observed
stellar parameters and evolutionary tracks.  As is the case for the SyXB star
V2116~Oph, the elemental
abundances do not show any peculiarities that would suggest either a previous
common envelope stage with the proto-NS massive star or that mass
was transferred during the SN event.  The velocity and orbit of V934~Her also appear
to be little affected by the SN suggesting it was an ECS.

The main sequence lifetime of the M giant progenitor was $\sim$2 Gyrs.  
The NS evolved from a massive star in Myrs, so the NS is nearly 
2 Gyrs old.  
Two other SyXB are known to be old, 4U 1954+319 \citep{enoto_et_al_2014}
and V2116~Oph/GX1+4 \citep{hinkle_et_al_2006}.  
Ages of Gyrs for the SyXB are similar to
ages derived for the NS in some of the standard LMXB 
\citep{wijnands_van_der_klis_1998}.
The observed NS properties are driven by mass 
accretion from the M giant stellar wind. In the case of 
V934~Her/4U~1700+24 neither X-ray nor radio 
pulsations have been detected from the NS component, 
but all evidence suggests this binary system is seen nearly pole on. 

We have compared the properties of the V934~Her M giant to those of the
M giant in the SyXB binary V2116 Oph.  Both are of similar
luminosity and both appear to be on the giant branch or early
AGB.  It seems likely that V2116 Oph is the most X-ray luminous of
the SyXB because of higher mass loss from its cooler M5/6 giant combined
with a relatively short, for an SyXB, 3.18 yr orbital period.  The
separation between the components in the V2116 Oph system is about
half that of the V934~Her system.  V2116 Oph and V934~Her are the
only two members of the SyXB group with determined orbits.  The
lack of optical emission lines in the M giant spectra and the 
ultra-long NS pulse periods in other SyXB strongly suggest that these systems 
are similar to V934~Her with long orbital periods.

\acknowledgments 

We are indebted to Duncan Galloway for sending his archival radial
velocity observations of V934~Her.  We thank Sharon Hunt for providing
several references critical to this project.  The FTS spectrum was
observed by Verne Smith.  We thank him for bringing the existence of
this spectrum to our attention.

This research was facilitated by the SIMBAD database, operated by
CDS in Strasbourg, France, and NASA's Astrophysics Data System
Abstract Service.  SM plot, by Robert Lupton and Patricia Monger,
was used in the production of some figures.  This work made use of
data from the European Space Agency (ESA) mission $Gaia$
(\url{https://www.cosmos.esa.int/gaia}), processed by the Gaia Data
Processing and Analysis Consortium (DPAC).  Funding for
the DPAC has been provided by national institutions, in particular
the institutions participating in the $Gaia$ Multilateral Agreement.

This research was based in part on
observations obtained at the Gemini Observatory, which is operated
by the Association of Universities for Research in Astronomy, Inc.,
under a cooperative agreement with the NSF on behalf of the Gemini
partnership: the National Science Foundation (United States), the
National Research Council (Canada), CONICYT (Chile), Ministerio de
Ciencia, Tecnolog\'{i}a e Innovaci\'{o}n Productiva (Argentina), and
Minist\'{e}rio da Ci\^{e}ncia, Tecnologia e Inova\c{c}\~{a}o (Brazil).  The Phoenix
spectrograph was developed by NOAO. IGRINS was developed under a
collaboration between the University of Texas at Austin and the
Korea Astronomy and Space Science Institute (KASI) with the financial
support of the US National Science Foundation under grant AST-1229522,
of the University of Texas at Austin, and of the Korean GMT Project
of KASI.  

The National Optical Astronomy Observatory is operated by the
Association of Universities for Research in Astronomy under cooperative
agreement with the National Science Foundation. KH and RJ express
their thanks to the Office of Science for support of their research.
The research at Tennessee State University was supported in part
by the State of Tennessee through its Centers of Excellence program.
JM has been financed by Polish National Science Centre (NSC) grants
2015/18/A/ST9/00746 and 2017/27/B/ST9/01940 and CG by NSC grant
SONATA No.~DEC-2015/19/D/ST9/02974.

\clearpage

\begin{figure}
\epsscale{0.9}
\plotone{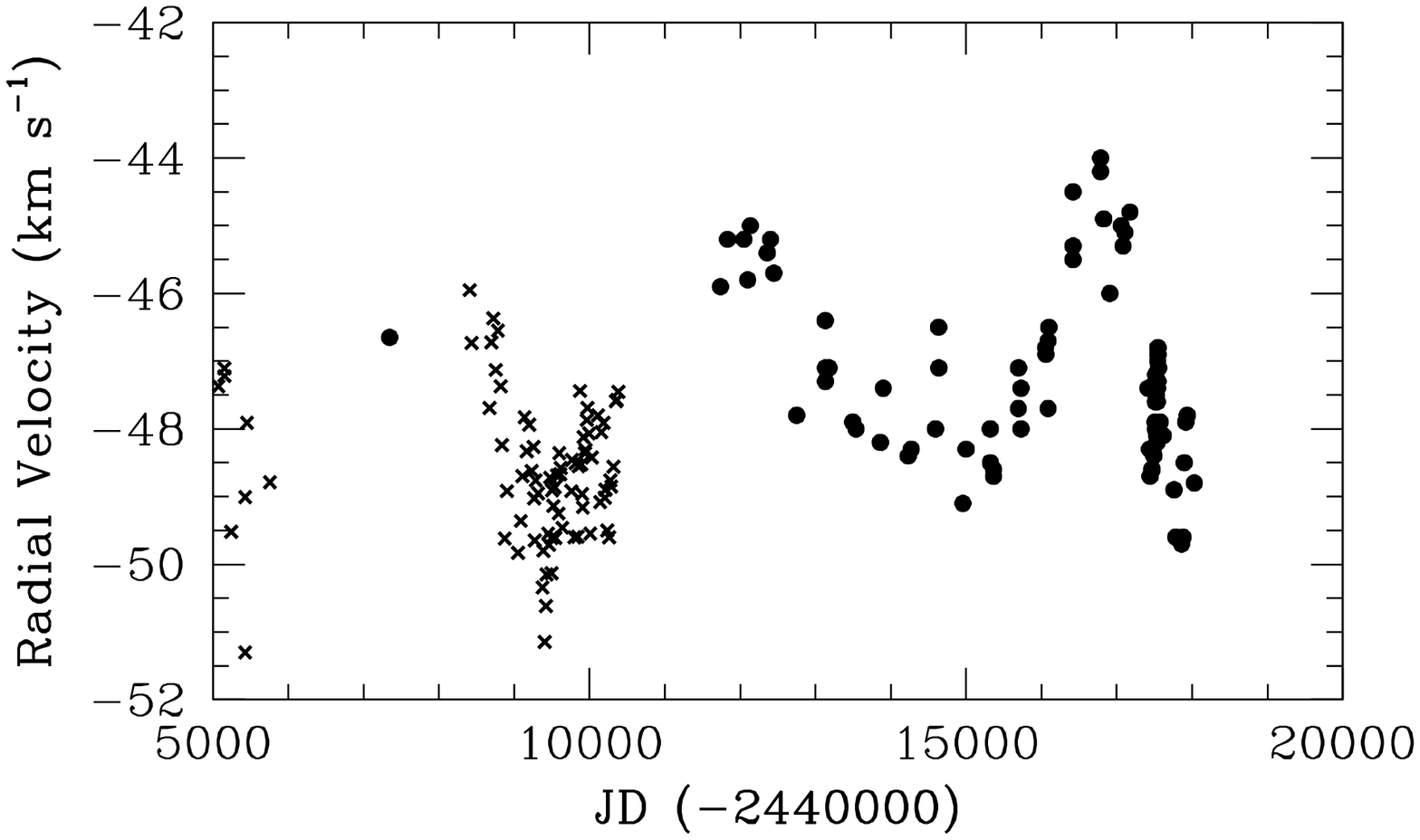}
\caption{
\label{f:vvst}
Radial velocity data (Table 1) for V934~Her as a function of time.
Solid circles = our velocities, x = Center of Astrophysics.
}
\end{figure}

\begin{figure} 
\epsscale{0.8} 
\plotone{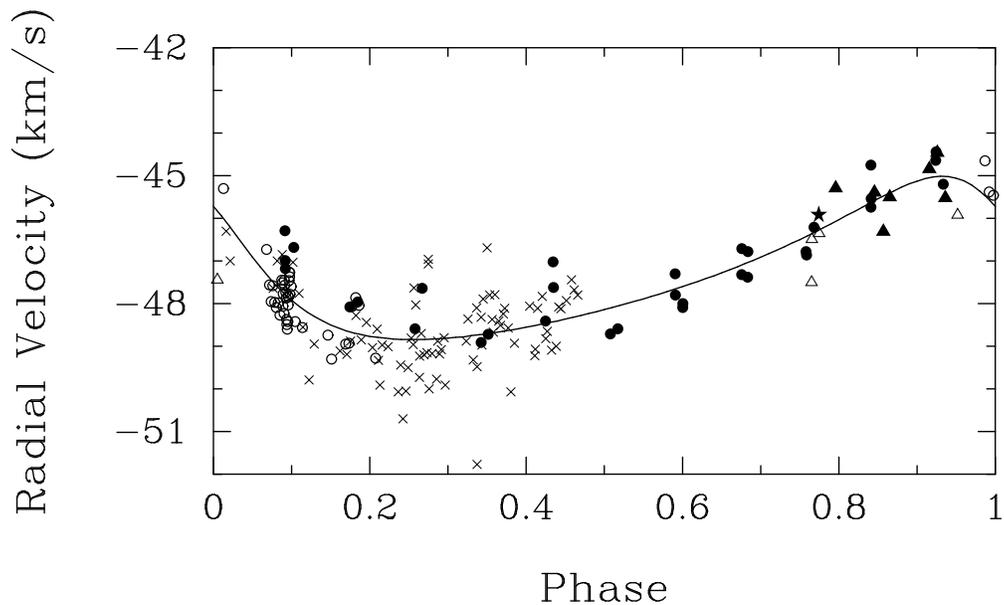}
\caption{
\label{f:orbit} 
The computed velocity curve of the 4391 day (12.0 yr) long-period orbit 
compared with
our radial velocities and those from CfA. Star = KPNO FTS, solid circle = 
KPNO coud\'e, open circle = Fairborn Observatory, solid triangle = 
Mount Stromlo Observatory and KPNO NICMASS, open triangle = KPNO and 
Gemini South Phoenix, x = Center for Astrophysics. Each plotted velocity 
consists of the total observed velocity minus its calculated short-period 
velocity. Zero phase is a time of periastron.
}
\end{figure}

\begin{figure} 
\epsscale{0.8} 
\plotone{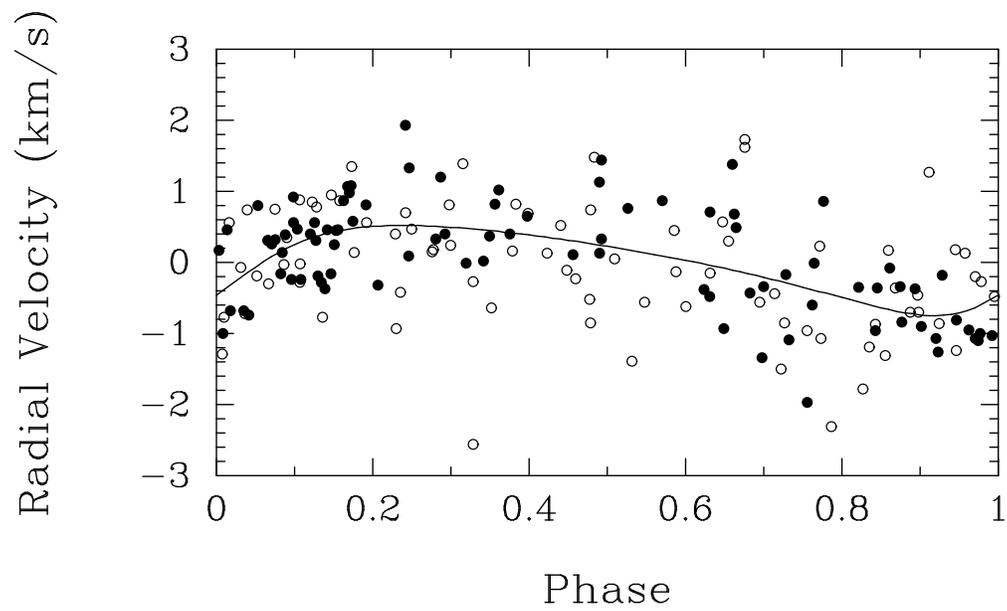}
\caption{
\label{f:res} 
The computed velocity curve of the 420.2 day velocity variation interpreted
as an orbit and compared with the velocity residuals. CfA velocities = 
open circles,
our velocities = solid circles. Each plotted velocity consists of the
total observed velocity minus its calculated long-period velocity.
Zero phase is a time of periastron.
}
\end{figure}

\begin{figure}
\epsscale{0.8}
\plotone{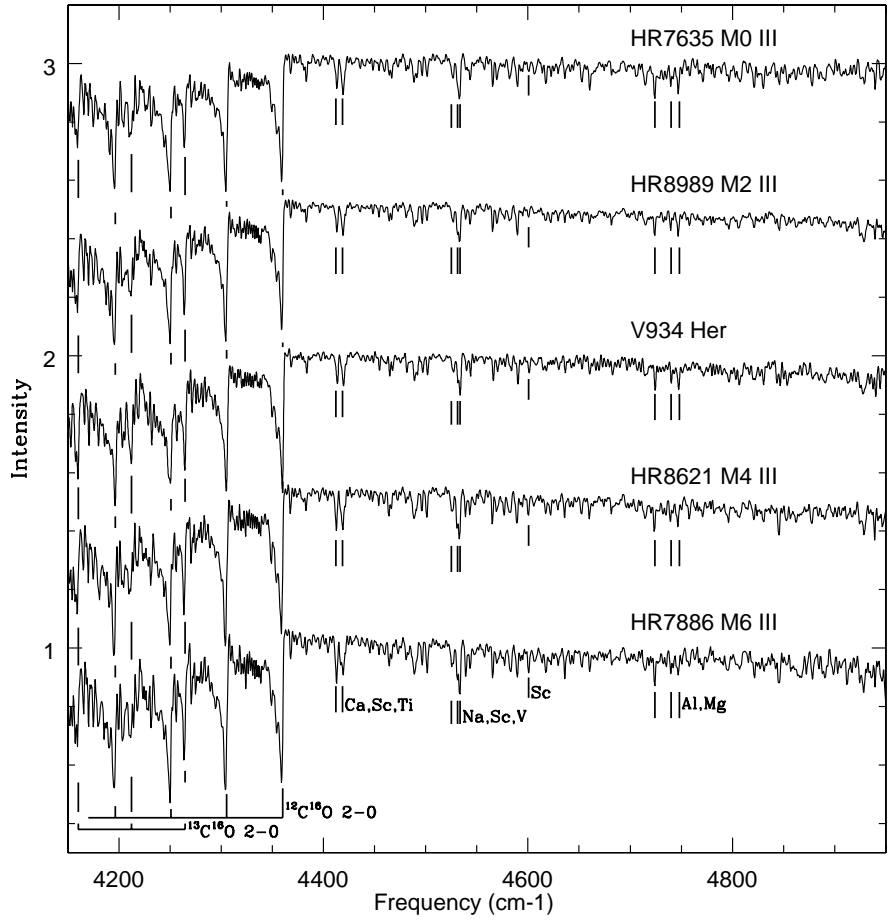}
\caption{
\label{f:K_band_low_res}
$K$-band FTS spectra at R$\sim$3000 comparing the spectrum of V934~Her to 
M0 III through M6 III standard star spectra from \citet{wallace_hinkle_1997}.
}
\end{figure}

\begin{figure}
\epsscale{0.8}
\plotone{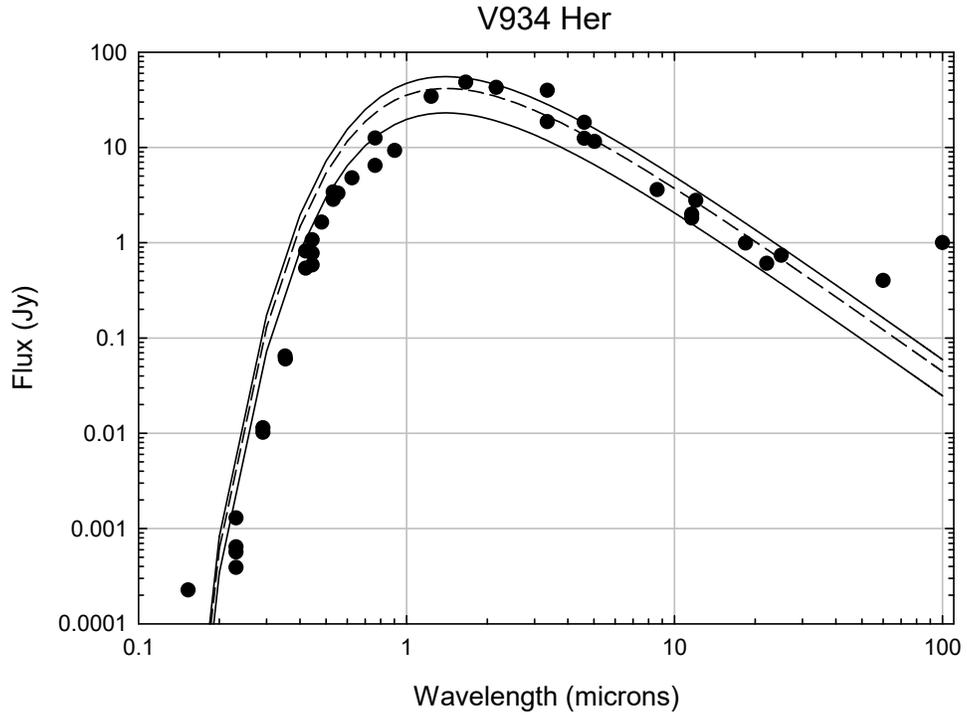}
\caption{
\label{f:flux_curve}
Photometry of V934~Her/4U~1700+24 from the literature compared to
a 3650 K blackbody.  The dashed line is a best fit while the two 
solid lines correspond to high and low envelopes.  
The blackbody integrated flux is 1.4 $\times$ 10$^{-10}$ W m$^{-2}$.  
Assuming the $Gaia$ distance of 544 pc the bolometric magnitude is $-$3.1.  
The blackbodies (lower
to upper fits) correspond to stellar radii of 70, 91, and 105~$R_\odot$.   
The blackbody fit shows IR excess suggesting a modest mass 
loss rate and UV excess due to the NS.  
}
\end{figure}

\begin{figure} 
\epsscale{0.98} 
\plotone{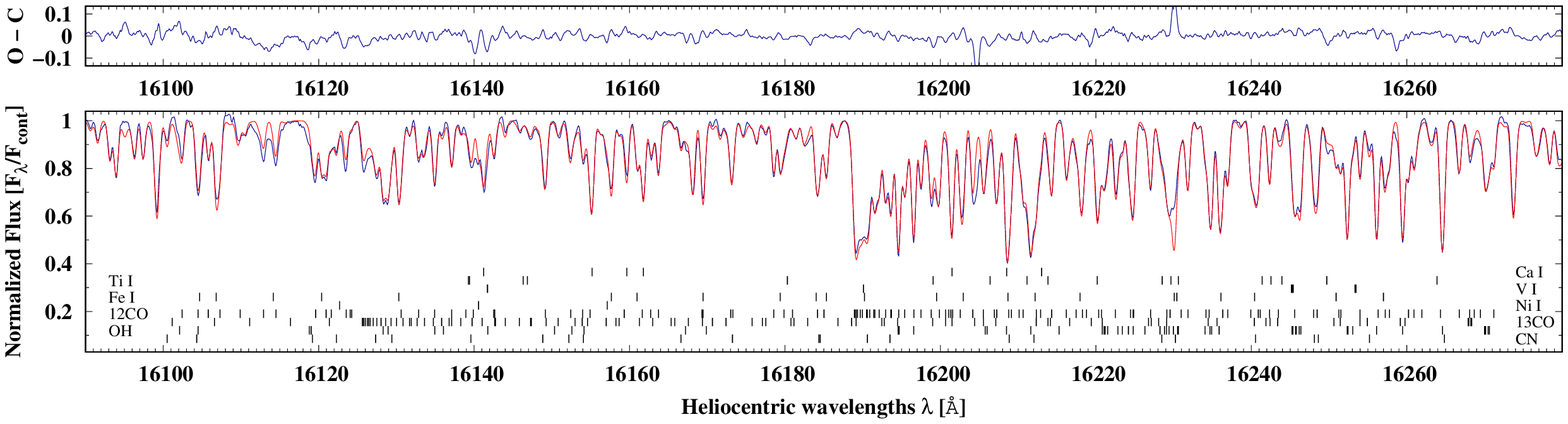}
\plotone{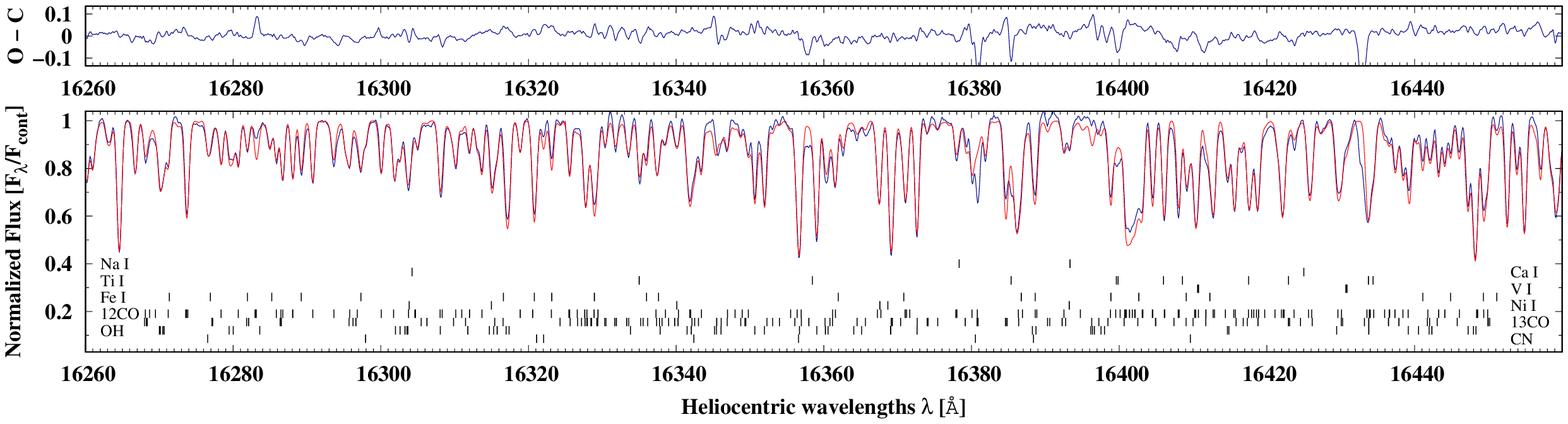}
\caption{
\label{f:IGRINS-H-2}
IGRINS $H$-band spectrum (blue line) shown as normalized flux as a
function of wavelength in \aa ngstr\"{o}ms in the region 16100--16440\,\AA\,
compared to the synthetic spectrum (red line) calculated with the
final abundances (Table~5).  Individual lines are identified with
dashes below the spectrum.  The $^{12}$C$^{16}$O 6-3 head is at
16190 \AA\ and the $^{12}$C$^{16}$O 7-4 band head is at 16400 \AA.
Residuals (O-C) to the fit are shown in the box above the spectrum.
} \end{figure}

\begin{figure} 
\epsscale{0.98} 
\plotone{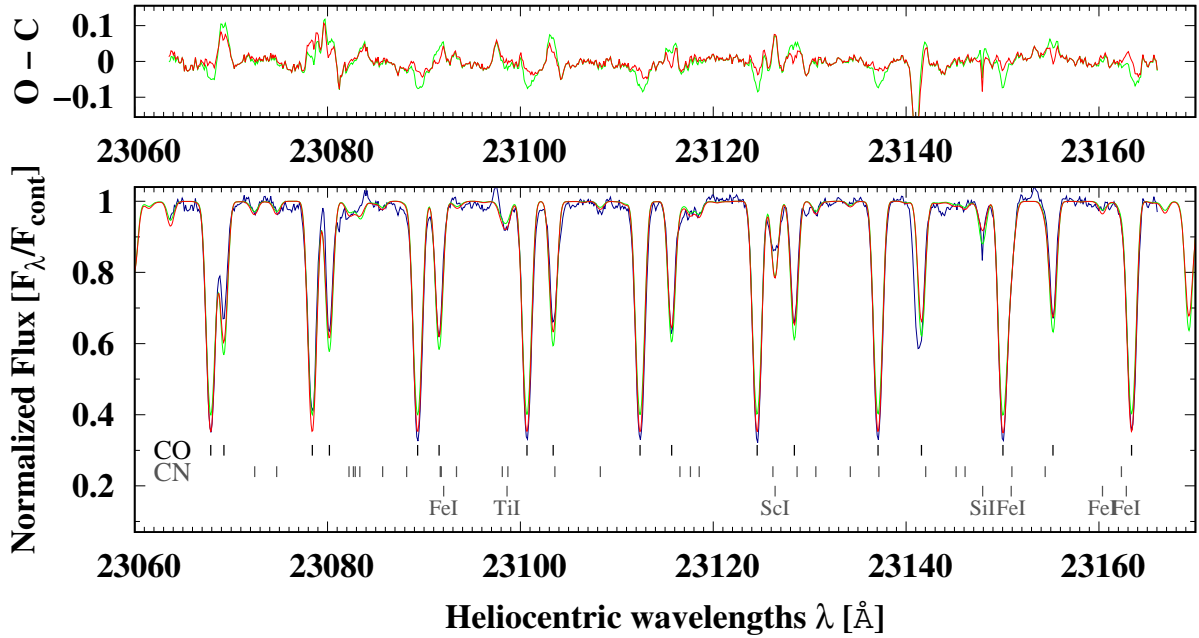}
\caption{
\label{f:Ph5} 
Lower box. Phoenix spectrum of V934\,Her observed 2012 June 8 (blue line) compared to
synthetic spectra obtained with two different adopted effective temperatures
$T_{\rm{eff}} = 3100$\,K (red line) and $3600$\,K (green line). The spectrum
is dominated by strong CO bands superimposed on a background of weak CN
and atomic lines. The hotter atmosphere does not give a good fit
to the alternating high and low excitation 2-0 $^{12}$C$^{16}$O lines.
Dashes below spectrum and upper box as in Figure 5.
}
\end{figure}

\begin{figure}
\plotone{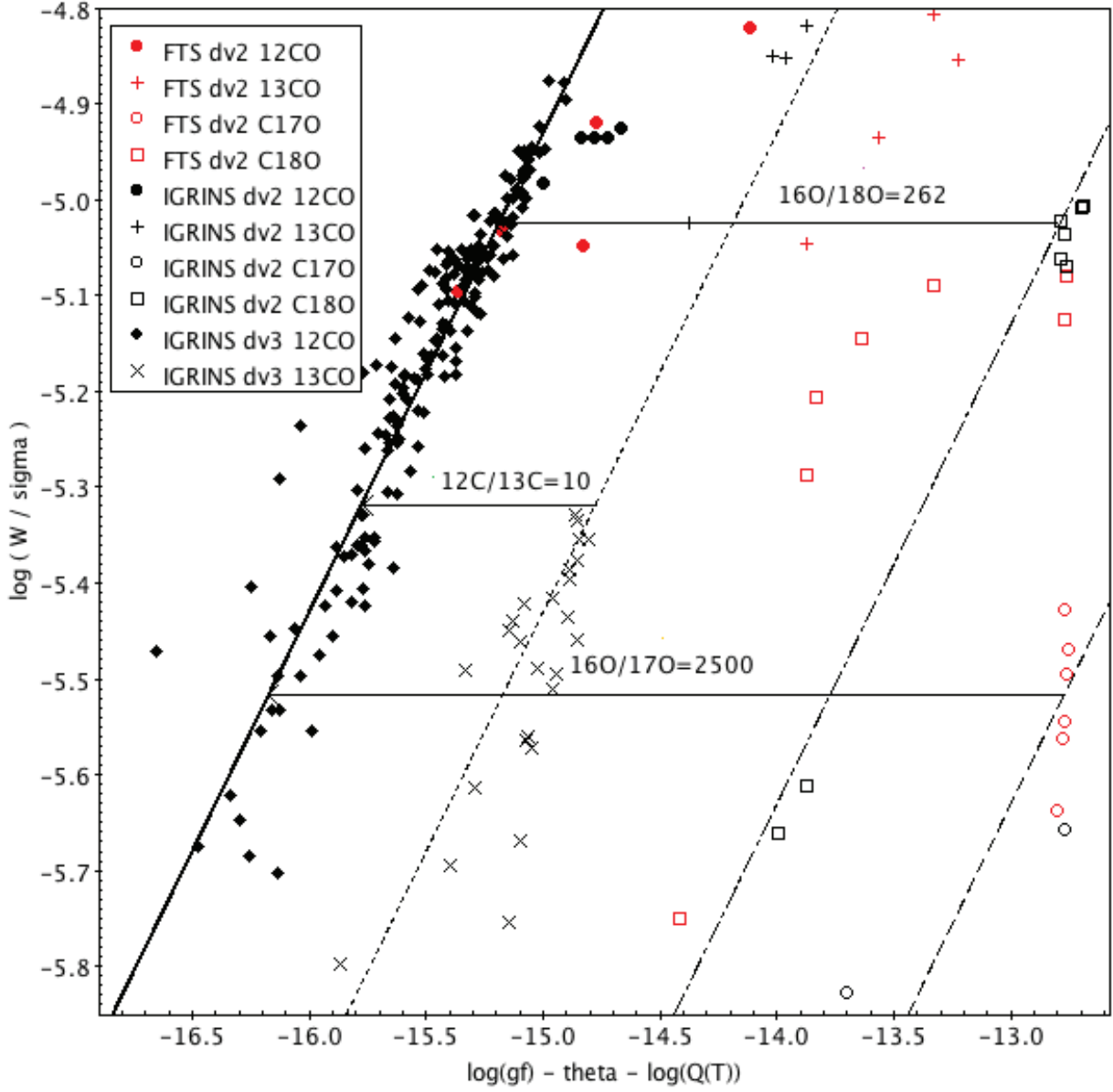}
\caption{
Curves-of-growth for the weak CO isotopologue lines in V934~Her.
Measurements from the IGRINS and FTS spectra are shown separately.
The IGRINS spectrum has a much higher S/N.   The IGRINS spectrum
covers the CO $\Delta$v=3 and $\Delta$v=2 regions while the FTS
spectrum only covers the CO $\Delta$v=2 region.  $\Delta$v=3 lines 
from the rare isotopologues were not detected.  The shifts between 
the curves-of-growth give the isotopic ratios (labelled). 
} 
\end{figure}

\clearpage

\begin{deluxetable}{lcrcccrl}
\tablenum{1}
\tablewidth{0pt}
\tablecaption{Radial Velocities of V934~Her}
\tablehead{\colhead{Helio. JD} & \colhead {RV} & \colhead{$O-C$} &
\colhead{} & \colhead{RV$_L$} & \colhead{} & 
\colhead{RV$_S$} & \colhead{} \\
\colhead{$-$ 2400000} & \colhead {km~s$^{-1}$} & \colhead{(km~s$^{-1}$)} &
\colhead{$\phi$$_L$} & \colhead{(km~s$^{-1}$)} & \colhead{$\phi$$_S$} 
& \colhead{(km~s$^{-1}$)} & \colhead{Source\tablenotemark{a}}
}
\startdata
 45072.3976 &  -47.37  &   1.22  &  0.256 &  -47.63  &  0.483  &   1.48 & CfA   \\
 45153.2525 &  -47.11  &   1.88  &  0.275 &  -46.96  &  0.676  &   1.73 & CfA   \\
 45153.2592 &  -47.22  &   1.77  &  0.275 &  -47.07  &  0.676  &   1.62 & CfA   \\
 45242.1427 &  -49.52  &   0.02  &  0.295 &  -48.80  &  0.887  &  -0.70 & CfA   \\
 45427.4203 &  -51.30  &  -3.03  &  0.337 &  -51.77  &  0.328  &  -2.56 & CfA   \\
 45427.4545 &  -49.01  &  -0.74  &  0.337 &  -49.48  &  0.328  &  -0.27 & CfA   \\
 45450.3925 &  -47.91  &   0.41  &  0.343 &  -48.32  &  0.383  &   0.82 & CfA   \\
 45754.5360 &  -48.79  &  -0.56  &  0.412 &  -49.07  &  0.107  &  -0.28 & CfA   \\
 47345.6300 &  -46.65  &   0.36  &  0.774 &  -45.92  &  0.894  &  -0.37 & KPNO1\tablenotemark{b} \\
 48408.3953 &  -45.95  &  -0.22  &  0.016 &  -46.30  &  0.423  &   0.13 & CfA  \\
 48431.2564 &  -46.73  &  -0.79  &  0.021 &  -47.00  &  0.477  &  -0.52 & CfA  \\
 48672.5247 &  -47.69  &  -0.14  &  0.076 &  -47.64  &  0.052  &  -0.19 & CfA  \\
 48695.5450 &  -46.72  &   0.60  &  0.082 &  -47.00  &  0.106  &   0.88 & CfA  \\
 48723.4965 &  -46.37  &   0.87  &  0.088 &  -46.85  &  0.173  &   1.35 & CfA  \\
 48752.4824 &  -47.13  &   0.18  &  0.095 &  -47.65  &  0.242  &   0.70 & CfA  \\
 48783.3251 &  -46.55  &   0.91  &  0.102 &  -47.03  &  0.315  &   1.39 & CfA  \\
 48818.2600 &  -47.37  &   0.30  &  0.110 &  -47.76  &  0.398  &   0.69 & CfA  \\
 48839.1402 &  -48.24  &  -0.43  &  0.114 &  -48.55  &  0.448  &  -0.11 & CfA  \\
 48874.1625 &  -49.62  &  -1.56  &  0.122 &  -49.79  &  0.532  &  -1.39 & CfA  \\
 48903.0909 &  -48.92  &  -0.65  &  0.129 &  -48.95  &  0.600  &  -0.62 & CfA  \\
 49048.4257 &  -49.83  &  -0.52  &  0.162 &  -49.11  &  0.946  &  -1.24 & CfA  \\
 49086.3421 &  -49.36  &  -0.55  &  0.171 &  -49.19  &  0.037  &  -0.72 & CfA  \\
 49107.3136 &  -48.70  &  -0.21  &  0.175 &  -48.88  &  0.086  &  -0.03 & CfA  \\
 49137.2635 &  -47.83  &   0.42  &  0.182 &  -48.28  &  0.158  &   0.87 & CfA  \\
 49167.1858 &  -48.33  &  -0.12  &  0.189 &  -48.85  &  0.229  &   0.40 & CfA  \\
 49196.1055 &  -47.94  &   0.32  &  0.196 &  -48.44  &  0.298  &   0.81 & CfA  \\
 49230.1082 &  -48.62  &  -0.26  &  0.203 &  -49.03  &  0.379  &   0.16 & CfA  \\
 49256.0466 &  -48.27  &   0.20  &  0.209 &  -48.60  &  0.440  &   0.52 & CfA  \\
 49263.9965 &  -49.03  &  -0.53  &  0.211 &  -49.33  &  0.459  &  -0.23 & CfA  \\
 49271.9952 &  -49.65  &  -1.11  &  0.213 &  -49.91  &  0.478  &  -0.85 & CfA  \\
 49284.9803 &  -48.76  &  -0.16  &  0.216 &  -48.97  &  0.509  &   0.05 & CfA  \\
 49317.9145 &  -48.95  &  -0.19  &  0.223 &  -49.01  &  0.588  &  -0.13 & CfA  \\
 49374.4494 &  -50.34  &  -1.23  &  0.236 &  -50.07  &  0.722  &  -1.50 & CfA  \\
 49388.4426 &  -49.80  &  -0.60  &  0.239 &  -49.44  &  0.756  &  -0.96 & CfA  \\
 49401.4404 &  -51.15  &  -1.85  &  0.242 &  -50.70  &  0.786  &  -2.31 & CfA  \\
 49418.4181 &  -50.62  &  -1.20  &  0.246 &  -50.05  &  0.827  &  -1.78 & CfA  \\
 49430.4187 &  -50.15  &  -0.65  &  0.249 &  -49.50  &  0.855  &  -1.31 & CfA  \\
 49448.3694 &  -49.55  &   0.03  &  0.253 &  -48.81  &  0.898  &  -0.70 & CfA  \\
 49459.3857 &  -49.71  &  -0.12  &  0.256 &  -48.96  &  0.924  &  -0.86 & CfA  \\
 49473.3215 &  -48.72  &   0.81  &  0.259 &  -48.03  &  0.958  &   0.13 & CfA  \\
 49494.2598 &  -50.13  &  -0.88  &  0.264 &  -49.73  &  0.007  &  -1.29 & CfA  \\
 49495.2301 &  -49.61  &  -0.38  &  0.264 &  -49.23  &  0.010  &  -0.77 & CfA  \\
 49504.3172 &  -48.91  &   0.15  &  0.266 &  -48.70  &  0.031  &  -0.07 & CfA  \\
 49519.2388 &  -49.14  &  -0.35  &  0.269 &  -49.20  &  0.067  &  -0.30 & CfA  \\
 49536.2015 &  -48.86  &  -0.31  &  0.273 &  -49.15  &  0.107  &  -0.02 & CfA  \\
 49548.1608 &  -49.61  &  -1.16  &  0.276 &  -50.00  &  0.136  &  -0.77 & CfA  \\
 49565.1973 &  -48.69  &  -0.34  &  0.280 &  -49.17  &  0.176  &   0.14 & CfA  \\
 49590.0435 &  -49.25  &  -0.94  &  0.285 &  -49.77  &  0.235  &  -0.42 & CfA  \\
 49596.0637 &  -48.36  &  -0.05  &  0.287 &  -48.88  &  0.250  &   0.47 & CfA  \\
 49607.0952 &  -48.67  &  -0.36  &  0.289 &  -49.18  &  0.276  &   0.15 & CfA  \\
 49617.0588 &  -48.58  &  -0.25  &  0.292 &  -49.08  &  0.300  &   0.24 & CfA  \\
 49638.9850 &  -49.46  &  -1.09  &  0.297 &  -49.91  &  0.352  &  -0.64 & CfA  \\
 49756.4591 &  -48.92  &  -0.11  &  0.323 &  -48.88  &  0.631  &  -0.15 & CfA  \\
 49766.3693 &  -48.46  &   0.40  &  0.326 &  -48.36  &  0.655  &   0.30 & CfA  \\
 49796.4229 &  -49.60  &  -0.57  &  0.332 &  -49.32  &  0.727  &  -0.85 & CfA  \\
 49815.3620 &  -48.51  &   0.64  &  0.337 &  -48.10  &  0.772  &   0.23 & CfA  \\
 49845.3133 &  -49.59  &  -0.25  &  0.344 &  -48.97  &  0.843  &  -0.87 & CfA  \\
 49852.2900 &  -48.55  &   0.83  &  0.345 &  -47.89  &  0.859  &   0.17 & CfA  \\
 49874.1502 &  -47.44  &   2.01  &  0.350 &  -46.69  &  0.911  &   1.27 & CfA  \\
 49888.2834 &  -48.52  &   0.90  &  0.353 &  -47.80  &  0.945  &   0.18 & CfA  \\
 49902.1412 &  -48.96  &   0.32  &  0.356 &  -48.37  &  0.978  &  -0.27 & CfA  \\
 49909.1852 &  -49.16  &   0.02  &  0.358 &  -48.67  &  0.995  &  -0.48 & CfA  \\
 49918.2127 &  -48.12  &   0.89  &  0.360 &  -47.79  &  0.016  &   0.56 & CfA  \\
 49939.1151 &  -48.36  &   0.25  &  0.365 &  -48.41  &  0.066  &   0.31 & CfA  \\
 49949.1086 &  -48.31  &   0.15  &  0.367 &  -48.51  &  0.090  &   0.35 & CfA  \\
 49965.1146 &  -47.87  &   0.41  &  0.371 &  -48.24  &  0.128  &   0.78 & CfA  \\
 49973.0348 &  -47.69  &   0.53  &  0.373 &  -48.11  &  0.147  &   0.95 & CfA  \\
 49992.0428 &  -48.07  &   0.06  &  0.377 &  -48.57  &  0.192  &   0.56 & CfA  \\
 50007.9674 &  -49.55  &  -1.45  &  0.381 &  -50.07  &  0.230  &  -0.93 & CfA  \\
 50027.9450 &  -48.42  &  -0.33  &  0.385 &  -48.93  &  0.278  &   0.18 & CfA  \\
 50112.4585 &  -47.80  &   0.47  &  0.404 &  -48.06  &  0.479  &   0.74 & CfA  \\
 50141.4002 &  -49.08  &  -0.70  &  0.411 &  -49.22  &  0.548  &  -0.56 & CfA  \\
 50157.4025 &  -48.05  &   0.39  &  0.415 &  -48.11  &  0.586  &   0.45 & CfA  \\
 50183.3229 &  -47.91  &   0.65  &  0.421 &  -47.83  &  0.647  &   0.57 & CfA  \\
 50203.2546 &  -49.02  &  -0.36  &  0.425 &  -48.82  &  0.695  &  -0.56 & CfA  \\
 50211.3116 &  -48.90  &  -0.20  &  0.427 &  -48.65  &  0.714  &  -0.44 & CfA  \\
 50236.2234 &  -49.50  &  -0.65  &  0.433 &  -49.09  &  0.773  &  -1.07 & CfA  \\
 50262.2075 &  -49.60  &  -0.59  &  0.438 &  -49.00  &  0.835  &  -1.19 & CfA  \\
 50276.1672 &  -48.76  &   0.32  &  0.442 &  -48.08  &  0.868  &  -0.36 & CfA  \\
 50288.1864 &  -48.85  &   0.27  &  0.444 &  -48.12  &  0.897  &  -0.46 & CfA  \\
 50319.1151 &  -48.56  &   0.43  &  0.451 &  -47.93  &  0.971  &  -0.20 & CfA  \\
 50348.0109 &  -47.59  &   0.89  &  0.458 &  -47.44  &  0.039  &   0.74 & CfA  \\
 50362.9830 &  -47.57  &   0.64  &  0.461 &  -47.68  &  0.075  &   0.75 & CfA  \\
 50382.9624 &  -47.45  &   0.50  &  0.466 &  -47.80  &  0.122  &   0.85 & CfA  \\
 51738.776  &  -45.90  &  -0.08  &  0.775 &  -46.35  &  0.349  &   0.37 & KPNO2\tablenotemark{c} \\
 51831.576  &  -45.20  &   0.78  &  0.796 &  -45.29  &  0.570  &   0.87 & KPNO3  \\
 52049.157  &  -45.20  &   0.20  &  0.845 &  -45.39  &  0.088  &   0.39 & MSO  \\
 52098.992  &  -45.80  &  -0.83  &  0.857 &  -46.31  &  0.207  &  -0.32 & MSO  \\
 52134.992  &  -45.00  &  -0.10  &  0.865 &  -45.50  &  0.292  &   0.40 & MSO  \\
 52357.318  &  -45.40  &   0.20  &  0.916 &  -44.84  &  0.821  &  -0.35 & MSO  \\
 52402.223  &  -45.20  &   0.56  &  0.926 &  -44.46  &  0.928  &  -0.18 & MSO  \\
 52447.045  &  -45.70  &  -0.50  &  0.936 &  -45.52  &  0.035  &  -0.68 & MSO  \\
 52749.825  &  -47.80  &  -1.61  &  0.005 &  -47.44  &  0.756  &  -1.97 & GemS\tablenotemark{d} \\
 53129.782  &  -46.40  &   1.49  &  0.092 &  -46.29  &  0.660  &   1.38 & KPNO4 \\
 53130.774  &  -47.10  &   0.80  &  0.092 &  -46.99  &  0.662  &   0.68 & KPNO4 \\
 53131.799  &  -47.30  &   0.61  &  0.092 &  -47.18  &  0.665  &   0.49 & KPNO4 \\
 53178.755  &  -47.10  &   1.29  &  0.103 &  -46.68  &  0.776  &   0.86 & KPNO4 \\
 53493.802  &  -47.90  &   0.58  &  0.175 &  -48.08  &  0.526  &   0.76 & KPNO4 \\
 53537.877  &  -48.00  &   0.75  &  0.185 &  -47.96  &  0.631  &   0.71 & KPNO4 \\
 53859.860  &  -48.20  &   0.26  &  0.258 &  -48.59  &  0.397  &   0.65 & KPNO4 \\
 53899.790  &  -47.40  &   1.20  &  0.267 &  -47.64  &  0.493  &   1.44 & KPNO4 \\
 54230.902  &  -48.40  &  -0.18  &  0.342 &  -48.91  &  0.281  &   0.33 & KPNO4 \\
 54270.753  &  -48.30  &  -0.02  &  0.351 &  -48.72  &  0.375  &   0.40 & KPNO4 \\
 54592.774  &  -48.00  &   0.05  &  0.425 &  -48.41  &  0.142  &   0.46 & KPNO4 \\
 54634.732  &  -46.50  &   1.41  &  0.434 &  -47.02  &  0.242  &   1.93 & KPNO4 \\
 54636.814  &  -47.10  &   0.81  &  0.435 &  -47.62  &  0.247  &   1.33 & KPNO4 \\
 54956.864  &  -49.10  &  -0.60  &  0.508 &  -48.71  &  0.008  &  -1.00 & KPNO4 \\
 54998.716  &  -48.30  &  -0.53  &  0.517 &  -48.59  &  0.108  &  -0.24 & KPNO4 \\
 55320.783  &  -48.00  &   0.35  &  0.591 &  -47.30  &  0.874  &  -0.34 & KPNO4 \\
 55321.715  &  -48.50  &  -0.14  &  0.591 &  -47.80  &  0.877  &  -0.84 & KPNO4 \\
 55362.751  &  -48.70  &  -0.49  &  0.600 &  -48.09  &  0.974  &  -1.10 & KPNO4 \\
 55363.758  &  -48.60  &  -0.40  &  0.600 &  -48.00  &  0.977  &  -1.00 & KPNO4 \\
 55693.744  &  -47.70  &  -0.22  &  0.676 &  -47.32  &  0.762  &  -0.60 & KPNO4 \\
 55694.722  &  -47.10  &   0.38  &  0.676 &  -46.71  &  0.764  &  -0.01 & KPNO4 \\
 55727.665  &  -48.00  &  -0.34  &  0.683 &  -47.38  &  0.843  &  -0.96 & KPNO4 \\
 55728.678  &  -47.40  &   0.26  &  0.684 &  -46.78  &  0.845  &  -0.36 & KPNO4 \\
 56055.748  &  -46.80  &  -0.35  &  0.758 &  -46.78  &  0.624  &  -0.38 & KPNO4 \\
 56058.771  &  -46.90  &  -0.44  &  0.759 &  -46.86  &  0.631  &  -0.48 & KPNO4 \\
 56086.865  &  -47.70  &  -1.14  &  0.765 &  -47.50  &  0.698  &  -1.34 & KPNO2\tablenotemark{e} \\
 56087.748  &  -46.70  &  -0.13  &  0.765 &  -46.49  &  0.700  &  -0.34 & KPNO2\tablenotemark{f}  \\
 56099.697  &  -46.50  &   0.12  &  0.768 &  -46.21  &  0.728  &  -0.17 & KPNO4 \\
 56419.693  &  -44.50  &   0.89  &  0.841 &  -44.75  &  0.490  &   1.13 & KPNO4 \\
 56419.766  &  -45.50  &  -0.11  &  0.841 &  -45.74  &  0.490  &   0.13 & KPNO4 \\
 56420.750  &  -45.30  &   0.09  &  0.841 &  -45.54  &  0.492  &   0.33 & KPNO4 \\
 56783.698  &  -44.20  &   0.38  &  0.924 &  -44.64  &  0.356  &   0.82 & KPNO4 \\
 56785.764  &  -44.00  &   0.59  &  0.924 &  -44.44  &  0.361  &   1.02 & KPNO4 \\
 56825.669  &  -44.90  &  -0.19  &  0.933 &  -45.20  &  0.456  &   0.11 & KPNO4 \\
 56906.677  &  -46.00  &  -0.85  &  0.952 &  -45.92  &  0.649  &  -0.93 & KPNO5\tablenotemark{g} \\
 57059.966  &  -45.00  &   0.82  &  0.987 &  -44.65  &  0.014  &   0.46 & Fair \\
 57083.951  &  -45.30  &   0.18  &  0.992 &  -45.38  &  0.071  &   0.26 & Fair \\
 57106.882  &  -45.10  &   0.21  &  0.997 &  -45.46  &  0.125  &   0.56 & Fair \\
 57174.700  &  -44.80  &   0.70  &  0.013 &  -45.30  &  0.287  &   1.20 & Fair \\
 57416.020  &  -47.40  &   0.58  &  0.068 &  -46.73  &  0.861  &  -0.08 & Fair \\
 57432.955  &  -48.30  &  -0.16  &  0.072 &  -47.56  &  0.901  &  -0.90 & Fair \\
 57442.031  &  -48.70  &  -0.51  &  0.074 &  -47.95  &  0.923  &  -1.26 & Fair \\
 57451.878  &  -48.30  &  -0.09  &  0.076 &  -47.58  &  0.947  &  -0.81 & Fair \\
 57462.001  &  -48.60  &  -0.43  &  0.078 &  -47.97  &  0.971  &  -1.07 & Fair \\
 57470.990  &  -48.60  &  -0.51  &  0.080 &  -48.09  &  0.992  &  -1.03 & Fair \\
 57481.825  &  -48.30  &  -0.36  &  0.083 &  -47.98  &  0.018  &  -0.68 & Fair \\
 57491.778  &  -48.40  &  -0.60  &  0.085 &  -48.27  &  0.041  &  -0.74 & Fair \\
 57501.968  &  -47.40  &   0.26  &  0.087 &  -47.45  &  0.066  &   0.31 & Fair \\
 57505.970  &  -47.40  &   0.21  &  0.088 &  -47.51  &  0.075  &   0.32 & Fair \\
 57508.976  &  -47.90  &  -0.32  &  0.089 &  -48.06  &  0.082  &  -0.16 & Fair \\
 57509.786  &  -47.60  &  -0.03  &  0.089 &  -47.77  &  0.084  &   0.14 & Fair \\
 57514.707  &  -48.00  &  -0.47  &  0.090 &  -48.23  &  0.096  &  -0.24 & Fair \\
 57515.845  &  -47.20  &   0.32  &  0.091 &  -47.45  &  0.099  &   0.56 & Fair \\
 57517.789  &  -47.30  &   0.20  &  0.091 &  -47.57  &  0.103  &   0.47 & Fair \\
 57524.813  &  -47.40  &   0.06  &  0.093 &  -47.74  &  0.120  &   0.40 & Fair \\
 57527.783  &  -47.50  &  -0.06  &  0.093 &  -47.87  &  0.127  &   0.31 & Fair \\
 57528.942  &  -48.00  &  -0.56  &  0.094 &  -48.37  &  0.130  &  -0.19 & Fair \\
 57530.737  &  -48.10  &  -0.67  &  0.094 &  -48.49  &  0.134  &  -0.28 & Fair \\
 57532.678  &  -48.20  &  -0.77  &  0.094 &  -48.60  &  0.139  &  -0.37 & Fair \\
 57535.865  &  -48.00  &  -0.58  &  0.095 &  -48.42  &  0.146  &  -0.16 & Fair \\
 57537.670  &  -47.60  &  -0.19  &  0.096 &  -48.03  &  0.151  &   0.25 & Fair \\
 57538.693  &  -47.40  &   0.01  &  0.096 &  -47.84  &  0.153  &   0.45 & Fair \\
 57539.670  &  -47.40  &   0.01  &  0.096 &  -47.84  &  0.155  &   0.46 & Fair \\
 57542.671  &  -47.00  &   0.41  &  0.097 &  -47.46  &  0.163  &   0.87 & Fair \\
 57544.808  &  -46.80  &   0.61  &  0.097 &  -47.27  &  0.168  &   1.07 & Fair \\
 57545.764  &  -46.90  &   0.51  &  0.097 &  -47.37  &  0.170  &   0.98 & Fair \\
 57546.725  &  -46.80  &   0.61  &  0.098 &  -47.27  &  0.172  &   1.08 & Fair \\
 57547.672  &  -47.30  &   0.11  &  0.098 &  -47.78  &  0.174  &   0.58 & Fair \\
 57554.783  &  -47.10  &   0.31  &  0.099 &  -47.60  &  0.191  &   0.81 & Fair \\
 57577.688  &  -47.90  &  -0.43  &  0.105 &  -48.42  &  0.246  &   0.09 & Fair \\
 57617.823  &  -48.10  &  -0.44  &  0.114 &  -48.56  &  0.341  &   0.02 & Fair \\
 57761.018  &  -48.90  &  -0.27  &  0.146 &  -48.74  &  0.682  &  -0.43 & Fair \\
 57781.997  &  -49.60  &  -0.79  &  0.151 &  -49.30  &  0.732  &  -1.09 & Fair \\
 57860.987  &  -49.70  &  -0.32  &  0.169 &  -48.95  &  0.920  &  -1.07 & Fair \\
 57878.716  &  -49.60  &  -0.28  &  0.173 &  -48.93  &  0.962  &  -0.95 & Fair \\
 57895.906  &  -48.50  &   0.61  &  0.177 &  -48.07  &  0.003  &   0.17 & Fair \\
 57916.838  &  -47.90  &   0.84  &  0.182 &  -47.86  &  0.053  &   0.80 & Fair \\
 57935.891  &  -47.80  &   0.67  &  0.186 &  -48.04  &  0.098  &   0.92 & Fair \\
 58028.686  &  -48.80  &  -0.49  &  0.207 &  -49.28  &  0.319  &  -0.01 & Fair \\
\enddata
\tablenotetext{a}{
CfA = Center for Astrophysics, KPNO1 = KPNO 4m + FTS,
KPNO2 = KPNO 2.1m + Phoenix, KPNO3 = KPNO coud\'e feed + NICMASS, MSO = 
Mount Stromlo Observatory 1.88m + NICMASS, GemS = Gemini South 8m + Phoenix,
KPNO4 = KPNO coud\'e feed + LB1A, KPNO5 = KPNO 4m + Phoenix, Fair = 
Fairborn Observatory
}  
\tablenotetext{b}{2.335 $\mu$m}
\tablenotetext{c}{1.557 $\mu$m}
\tablenotetext{d}{2.226 $\mu$m, R = 70000}
\tablenotetext{e}{2.311 $\mu$m}
\tablenotetext{f}{1.562 $\mu$m}
\tablenotetext{g}{1.563 $\mu$m}
\end{deluxetable}

\begin{deluxetable}{lllclrr}
\tablenum{2}
\tablewidth{0pt}
\tablecaption{Spectra used for abundance analysis}
\tablehead{\colhead{ID} &\colhead{Date} & \colhead{Helio. JD} 
& \colhead{Spec. Region} & \colhead{Instrument} & \colhead{Res.} & \colhead{S/N} \\ 
           \colhead{}           &\colhead{(UT)} & \colhead{}    
& \colhead{(\AA)}           & \colhead{}           & \colhead{($\lambda/\Delta\lambda$)}     & \colhead{}} 
\startdata
Ph1    & 2000 Jul 13 & 2451738.77 & 15590 -- 15662 & Phx/KPNO 2.1 & 50000  & $\sim$100  \\ 
Ph2    & 2012 Jun 09 & 2456087.75 & 15590 -- 15655 & Phx/KPNO 2.1 & 50000  & $\sim$100  \\ 
Ph3    & 2014 Sep 06 & 2456906.68 & 15600 -- 15665 & Phx/KPNO 4   & 50000  & $\sim$100  \\ 
Ph4    & 2003 Apr 20 & 2452749.82 & 22214 -- 22320 & Phx/GS       & 70000  & $\sim$100  \\ 
Ph5    & 2012 Jun 08 & 2456086.86 & 23060 -- 23162 & Phx/KPNO 2.1 & 50000  & $\sim$100  \\ 
FTS    & 1988 Jul 03 & 2447345.5~ & 20800 -- 24050 & KPNO FTS 4   & 32000  & 63         \\ 
IGRINS & 2018 Apr 22 & 2458230.84 & 15000 -- 17000 & IGRINS/GS    & 45000  & $>$100     \\ 
IGRINS & 2018 Apr 22 & 2458230.84 & 20800 -- 24050 & IGRINS/GS    & 45000  & $>$100     \\ 
\enddata
\end{deluxetable}

\begin{deluxetable}{lcc}
\tablenum{3}
\tablewidth{0pt}
\tablecaption{Orbital Elements and Related Parameters of V934~Her}

\tablehead{\colhead{Parameter} & \colhead{LSP} & \colhead{Orbit} 
}
\startdata
$P$ (days)             & 420.17 $\pm$ 0.79    & 4391 $\pm$ 33 \\
$P$ (yr)               & 1.150 $\pm$ 0.002    & 12.02 $\pm$ 0.09 \\
$T$ (HJD)              & 2457894 $\pm$ 22     & 2457118 $\pm$ 89  \\
$\gamma$ (km~s$^{-1}$) & ...                  & $-$47.358 $\pm$ 0.063 \\
$K$ (km~s$^{-1}$)      & 0.634  $\pm$ 0.080 & 1.915 $\pm$ 0.097 \\
$e$                    & 0.33   $\pm$ 0.11  & 0.354 $\pm$ 0.036 \\
$\omega$ (deg)         & 237    $\pm$ 23   & 50.7 $\pm$  8.8 \\
$a$~sin~$i$ (10$^6$ km) & 3.45  $\pm$ 0.52  & 108.2 $\pm$ 6.6    \\
$f(m)$ ($M_{\sun}$)    & 0.0000093 $\pm$ 0.0000042 & 0.00217 $\pm$ 0.00047 \\
Standard error of an observation &     &     \\
of unit weight (km~s$^{-1}$) & 0.6 & 0.6 \\
\enddata
\end{deluxetable}

\begin{deluxetable}{lcl}
\tablenum{4}
\tablewidth{0pt}
\tablecaption{Parameters of the V934~Her M III }

\tablehead{\colhead{Parameter} & \colhead{Value} & \colhead{Source} 
}
\startdata
Distance              & 544 $\pm$ 10 pc             & $Gaia$                                 \\
Spec Type             & M3 III                     & Fig. 3; T$_{eff}$ \& luminosity      \\
$T_{\rm{eff}}$       & 3650 $\pm$ 100 K            & Sp.Ty.; V-K; CO T$_{exc}$            \\
Luminosity            & 1200 $\pm$ 200 L$_\odot$   & See text; Fig. 4                      \\
Radius                & 90 $\pm$ 20 R$_\odot$       & Fig. 4; \citet{van_belle_et_al_1999} \\
Mass                  & 1.6 $^{+0.1}_{-0.2}$ M$_\odot$  & Evol. tracks \& mass loss            \\
Surface gravity (log\,g) & 0.7 $\pm$ 0.2 (cm s$^{-1}$) & Mass and radius                      \\
Inclination\tablenotemark{a} & 11$\,.\!\!^\circ$3$\pm$0$\,.\!\!^\circ$4 & Assume equator and orbit coplanar \\
$[Fe/H]$              &   -0.60$\pm$0.10            & See text                             \\
$[\alpha/H]$          &   -0.33$\pm$0.12             & See text (Mg+Si+Ca)                  \\
Age                   &   $\sim$ 2 Gyrs           & Evol. tracks        \\
\enddata
\tablenotetext{a}{Equator to plane of sky}
\end{deluxetable}

\begin{deluxetable}{ccccccc}
\tablenum{5}
\tablewidth{0pt}
\tablecaption{Abundance Summary} 
\tablehead{ \colhead{Element} &       \multicolumn2c{FTS}                &       \multicolumn2c{Phoenix}   & \multicolumn2c{IGRINS\tablenotemark{a}}  \\
           \colhead{} &  \colhead{$\log{\epsilon(X)}$\tablenotemark{b}} & \colhead{[$X$]\tablenotemark{c}} & 
\colhead{$\log{\epsilon(X)}$\tablenotemark{b}} & \colhead{[$X$]\tablenotemark{c}} & \colhead{$\log{\epsilon(X)}$\tablenotemark{b}} & \colhead{[$X$]\tablenotemark{c}} 
}
\startdata
C                      &  $7.79\pm0.05$ & $-0.64\pm0.10$ & $7.91\pm0.03$ & $-0.52\pm0.08$ & $7.84\pm0.01$ & $-0.59\pm0.06$ \\
N                      &  $8.23\pm0.11$ & $+0.40\pm0.16$ & $7.69\pm0.06$ & $-0.14\pm0.11$ & $7.63\pm0.03$ & $-0.20\pm0.08$  \\
O                      &  $8.27\pm0.07$ & $-0.42\pm0.12$ & $8.35\pm0.03$ & $-0.34\pm0.08$ & $8.29\pm0.01$ & $-0.40\pm0.06$  \\
Na                     &  $5.72\pm0.21$ & $-0.49\pm0.25$ & \nodata       & \nodata        & $5.44\pm0.11$ & $-0.77\pm0.15$  \\ 
Mg                     &  \nodata      & \nodata       & \nodata      & \nodata       & $7.35\pm0.03$ & $-0.24\pm0.07$  \\ 
Al                     &  $6.45\pm0.18$ & $+0.02\pm0.22$ & \nodata & \nodata             & $6.08\pm0.08$ & $-0.35\pm0.12$  \\ 
Si                     &  $7.71\pm0.13$ & $+0.20\pm0.16$ & \nodata & \nodata            & $7.23\pm0.05$ & $-0.28\pm0.08$  \\ 
S                      &  $7.10\pm0.27$ & $-0.02\pm0.30$ & \nodata & \nodata             & \nodata       & \nodata         \\
K                      & \nodata        & \nodata        &  \nodata &  \nodata           & $4.75\pm0.08$ & $-0.29\pm0.13$  \\ 
Ca                     &  $6.07\pm0.13$ & $-0.25\pm0.16$ & \nodata & \nodata             & $5.84\pm0.05$ & $-0.48\pm0.08$  \\ 
Sc                     &  $3.17\pm0.10$ & $+0.01\pm0.14$ & $2.96\pm0.21$ & $-0.20\pm0.25$  & \nodata       & \nodata         \\
Ti                     &  $4.76\pm0.11$ & $-0.17\pm0.15$ & $4.52\pm0.15$ & $-0.41\pm0.19$  & $4.43\pm0.06$ & $-0.50\pm0.10$  \\ 
V                      & \nodata        & \nodata        &  \nodata &  \nodata            & $3.32\pm0.11$ & $-0.57\pm0.19$  \\ 
Cr                     & \nodata        & \nodata        &  \nodata &  \nodata  & $4.94\pm0.07$ & $-0.68\pm0.11$  \\ 
Mn                     & \nodata        & \nodata        &  \nodata &  \nodata  & $4.84\pm0.11$ & $-0.58\pm0.15$  \\ 
Fe                     &  $7.03\pm0.08$ & $-0.44\pm0.12$ & $6.93\pm0.07$ & $-0.54\pm0.11$  & $6.87\pm0.01$ & $-0.60\pm0.05$  \\ 
Co                     & \nodata        & \nodata        & \nodata  & \nodata   & $4.37\pm0.05$ & $-0.56\pm0.10$  \\ 
Ni                     &  $6.50\pm0.29$ & $+0.30\pm0.33$ & $5.85\pm0.15$ & $-0.35\pm0.19$  & $5.66\pm0.08$ & $-0.54\pm0.12$  \\ 
%
$^{12}$C/$^{13}$C      &  \multicolumn2c{$6.7\pm0.9$}  & \multicolumn2c{\nodata}  & \multicolumn2c{$10.2\pm0.3$}    \\
$^{16}$O/$^{17}$O      &  \multicolumn2c{$4500\pm700$} & \multicolumn2c{\nodata}  & \multicolumn2c{$2750\pm320$\tablenotemark{d}}      \\
$^{16}$O/$^{18}$O      &  \multicolumn2c{$390\pm90$}   & \multicolumn2c{\nodata}  & \multicolumn2c{$250\pm30$\tablenotemark{d}} \\
\enddata
\tablenotetext{a}{From $H$ band spectrum}
\tablenotetext{b}{ $\log{\epsilon}(X) = \log{(N(X) N(H)^{-1})} + 12.0$.
Uncertainty is \small{3$\sigma$} from the fit. See Table 7 for the total uncertainty.  Abundances in dex.}
\tablenotetext{c}{\small{Relative to the Sun [$X$] abundances in respect to the solar composition of
\citet{Asp2009}, \citet{Sco2015a} and \citet{Sco2015b}}}
\tablenotetext{d}{From $K$ band spectrum}
\end{deluxetable}

\clearpage

\begin{deluxetable}{ccc}
\tablenum{6}
\tablewidth{0pt}
\tablecaption{Line Broadening Parameters}
\tablehead{     \colhead{Spectrum} & \colhead{$\zeta_{\rm t}$ (km s$^{-1}$)} & \colhead{$\xi_{\rm t}$ (km s$^{-1}$)} }
\startdata
FTS    &  $5.84\pm0.74$  & $2.49\pm0.09$  \\
Ph1    &  $3.65\pm0.65$  & $2.47\pm0.12$  \\
Ph2    &  $4.49\pm0.59$  & $2.47\pm0.12$  \\
Ph3    &  $4.16\pm0.51$  & $2.47\pm0.12$  \\
Ph4    &  $6.67\pm0.35$  & $2.47\pm0.12$  \\
Ph5    &  $5.14\pm0.52$  & $2.47\pm0.12$  \\
IGRINS &  $5.41\pm0.34$  & $2.31\pm0.07$  \\
\enddata
\end{deluxetable}

\clearpage

\begin{deluxetable}{ccccc}
\tablenum{7}
\tablewidth{0pt}
\tablecaption{Abundance Uncertainty Summary\tablenotemark{a}}
\tablehead{\colhead{Element} & \colhead{$\Delta T_{\rm{eff}} = +100$\,K} & \colhead{$\Delta \log{g} = +0.5$} & \colhead{$\Delta  \xi_{\rm{t}} = +0.1$} & \colhead{$\Delta$\tablenotemark{b}}
}
\startdata
C   & $+0.04$ & $+0.19$ & ~$0.00$ & $\pm0.20$ \\
N   & $+0.05$ & $-0.06$ & $+0.01$ & $\pm0.08$ \\
O   & $+0.12$ & $+0.03$ & $-0.01$ & $\pm0.13$ \\
Na  & $+0.13$ & $-0.20$ & $-0.02$ & $\pm0.24$ \\
Mg  & $+0.03$ & $-0.09$ & $+0.03$ & $\pm0.10$ \\
Al  & $+0.06$ & $-0.03$ & $-0.03$ & $\pm0.08$ \\
Si  & $-0.04$ & $+0.11$ & $+0.02$ & $\pm0.12$ \\
K   & $+0.03$ & $+0.04$ & $-0.01$ & $\pm0.06$ \\
Ca  & $+0.05$ & $+0.03$ & $-0.01$ & $\pm0.06$ \\
Ti  & $+0.07$ & $+0.09$ & $-0.02$ & $\pm0.12$ \\
V   & $+0.07$ & $+0.09$ & $-0.02$ & $\pm0.12$ \\
Cr  & $+0.05$ & $+0.05$ & $-0.01$ & $\pm0.08$ \\
Mn  & $-0.01$ & $+0.09$ & ~$0.00$ & $\pm0.10$ \\
Fe  & $-0.02$ & $+0.09$ & $-0.01$ & $\pm0.10$ \\
Co  & $+0.01$ & $+0.12$ & $+0.01$ & $\pm0.13$ \\
Ni  & $-0.02$ & $+0.11$ & ~$0.00$ & $\pm0.12$ \\
\enddata
\tablenotetext{a}{$H$-band~IGRINS~data~and~model~atmospheres~with
parameters $T_{\rm{eff}} = 3600$\,K, $\log{g} = 0.5$, and $\xi_{\rm{t}} = +2.3$.}
\tablenotetext{b}{$[(\Delta T_{\rm{eff}})^2 + (\Delta \log{g})^2 + (\Delta \xi_{\rm{t}})^2]^{0.5}$}
\end{deluxetable}

\end{document}